\documentclass[journal]{IEEEtran}

\usepackage{amsmath}
\usepackage{amssymb}
\usepackage{bm}
\usepackage{graphicx}
\usepackage{booktabs}
\usepackage{array}
\usepackage{float}
\usepackage{flushend}
\usepackage{xcolor}
\usepackage[numbers,sort&compress]{natbib}
\usepackage[colorlinks=true, citecolor=blue, urlcolor=blue,
            linkcolor=blue!60!black]{hyperref}%
\usepackage{fancyhdr}
\fancypagestyle{firstpage}{%
  \fancyhf{}%
  \fancyfoot[R]{\footnotesize FERMILAB-PUB-26-0726-SQMS}%
}

\makeatletter
\def\paragraph{\@startsection{paragraph}{4}{2\parindent}%
{0.6\baselineskip plus 0.2ex minus 0.1ex}%
{0ex}{\normalfont\normalsize\itshape}}%
\makeatother

\newcommand{\E}{\mathbb{E}}
\newcommand{\conj}[1]{{#1}^*}

\newcommand{\Var}{\operatorname{Var}}

\begin{document}

\title{Real-Time Adaptive Filtering and the Boxcar Limit in Superconducting Qubit Readout}

\author{
	Hans~Johnson,$^{1,2,\dagger}$ Tanay~Roy,$^{2}$ Leonardo~Bove,$^{2}$\\
	David~van~Zanten,$^{2}$ Silvia~Zorzetti,$^{2}$ and Jafar~Saniie$^{1}$\\[6pt]

{
	\small\itshape 
	$^{1}$Embedded Computing and Signal Processing (ECASP) Research Laboratory,\\
	Department of Electrical and Computer Engineering, Illinois Institute of Technology, Chicago, IL 60616, USA
}
\\[2pt]
{
	\small\itshape 
	$^{2}$Superconducting Quantum Materials and Systems (SQMS) Center,\\
	Fermi National Accelerator Laboratory, Batavia, IL 60510, USA
}%

\vspace{-4.0ex}%

\thanks{$^{\dagger}$\,\href{mailto:hjohnson@fnal.gov}{hjohnson@fnal.gov}}}

\maketitle
\thispagestyle{firstpage}

\begin{abstract}
Control and readout systems for superconducting quantum computing support real-time control and high-fidelity qubit-state discrimination. We examine when additional digital processing improves readout fidelity. A common baseline for dispersive readout uses a boxcar averager to compute a uniformly weighted average over a fixed window followed by threshold-based state assignment. We evaluated an adaptive finite-impulse-response (FIR) filter trained in real time using the least-mean-squares (LMS) algorithm on a field-programmable gate array (FPGA) quantum controller. We also analyzed measured readout shots offline to compare the frozen filter with boxcar averaging, and evaluate per-sample weighting plus sequential detection.
For the fixed FIR filters tested, summing the filtered samples gives the boxcar sum multiplied by one complex number apart from a small edge correction. We call this the kernel collapse. The multiplication scales the state separation and noise equally without improving state discrimination, and we call this the boxcar limit.
The frozen LMS filter reduces trace noise, but shows no resolved improvement in readout fidelity. On a 2D transmon, optimizing the readout window accounted for much of the improvement over full-window integration. We then examined how changing the decision method could reduce calibration effort and decision time. On a higher-fidelity 3D ancilla readout, our modeled template estimation reached the same fidelity target as direct averaging with 89.8\% fewer labeled shots per state. Offline sequential replay assigned most shots before the acquisition window ended. These results distinguish waveform denoising from state-discrimination improvement, and motivate direct control of decision weights and assignment time in dispersive readout.

\end{abstract}


\section{Introduction}
\label{sec:intro}

Interfacing with a quantum computer is a classical measurement instrumentation problem despite the quantum mechanical nature of quantum computation. The quantum bit (qubit) is the fundamental computational unit in a quantum computer. Quantum computation uses control operations to manipulate qubit states, and a separate set of readout operations to determine the qubit's state.
In dispersive readout~\cite{Krantz2019}, a probe tone interrogates a resonator coupled to the qubit rather than driving the qubit directly. The reflected or transmitted response of the resonator depends on the qubit state. The amplified response is digitized and reduced to a state assignment. 
Relaxation and dephasing, characterized by $T_1$ and $T_2$, constrain the timing and accuracy of operations that use this assignment~\cite{Krantz2019}. Quantum computing controllers are typically composed of digital electronics that interface with analog electronics. This lets researchers utilize software-defined radio (SDR) techniques for control and readout operations. Because these controllers sit at the border of digital and analog electronics, sophisticated digital signal processing methods must be employed to ensure accurate results~\cite{Walter2017,Krantz2019}. Regarding readout, there are many active avenues of research being conducted to refine how measurements are interpreted as well as real-time feedback during a live experiment, or "in situ"~\cite{Gambetta2007,Gambetta2008,Blais2021,Krantz2019}. These operations also incur processing and control latency. 

Commercial readout workflows typically use host-computed, frozen per-sample integration weights with hardware thresholding~\cite{ZurichInstruments2026,QuantumMachines2026,Qblox2026}.
Previous research on the readout chain includes machine-learning discrimination of measurement trajectories~\cite{Magesan2015} and optimal linear weights estimated from averaged trajectories~\cite{Bultink2018}. These integration weights are typically computed from labeled shots on a host and held fixed during readout.
Hidden-Markov posterior inference has also been applied to dispersive readout in host-side post-processing~\cite{Martinez2020}.
Other real-time approaches adapt the readout duration or the subsequent control rather than the readout statistic itself. Adaptive-decision speedups stop the acquisition once a running confidence reaches its target in a formulation that covers dispersive readout~\cite{DAnjou2016}. FPGA-resident Bayesian estimation uses thresholded qubit outcomes to update a relaxation-rate posterior and select the next waiting time~\cite{Berritta2026}. QubiCML on the QubiC FPGA-based stack~\cite{Vora2024QubiCML} combines offline-estimated per-sample weights with a compact host-trained network, and it cites this project's adaptive LMS proof of concept~\cite{Johnson2024}. Additionally, a co-designed neural-network workflow runs on QICK~\cite{DiGuglielmo2025,Stefanazzi2022}, which is the native FPGA controller used in this research.

The methods above either hold the readout statistic fixed during operation or adapt the acquisition and control around it. A simple baseline is boxcar integration, which weights all samples in the selected window equally. This paper asks when adaptive or sequential methods improve on that baseline and how architecture and operating-point statistics limit those gains. The research began from a simple conjecture: ensemble-averaged readout trajectories, the average of the same time sample across many shots, look clean while single shots appear noise-dominated. We expected an on-fabric filter trained on the average per-state waveform to improve single-shot fidelity, reduce the shots needed to estimate that fidelity, and simplify calibration. Against full-window integration and an optimized boxcar averager whose window start and length were chosen on training data, the filter showed no systematic fidelity gain at the measured operating points. For a given dataset, model class, and objective, moving optimization from the host to the FPGA does not by itself raise the best attainable fidelity. Its potential operational value lies in exposing convergence during calibration, holding thresholds stable across recalibrations, and responding to drift during operation. The systems above are therefore architectural references rather than fidelity comparators. This work builds and verifies an instrument that adapts in place, reports its implementation cost, and leaves the qubit-control path unchanged. The contribution is a measured operating-point record of gains and failure regimes for the tested readout methods, which guides the design and calibration of qubit readout electronics~\cite{Stefanazzi2022,Ryan2017,Vora2024QubiCML,DiGuglielmo2025}.

Sec.~\ref{sec:governing} derives the architectural limit and examines the available matched-filter and whitening headroom. Sec.~\ref{sec:instrument} describes the implemented filter and verifies its trace-noise reduction. Sec.~\ref{sec:map} reports the measured fidelity comparisons. Sec.~\ref{sec:calibration} evaluates model-based template calibration and offline sequential detection on the higher-fidelity ancilla readout. Future work will test an FPGA-based decision path in situ for per-shot state assignment and earlier active-reset feedback, and drift-triggered recalibration of templates and thresholds initiated by the adaptive filter. Sec.~\ref{sec:conclusion} concludes the work.

\section{The Governing Law: Kernel Collapse and the Matched-Filter Ceiling}
\label{sec:governing}

The governing result of this work is architectural. The deployed readout path applies a frozen, shift-invariant finite-impulse-response (FIR) filter and then accumulates every filtered sample with equal weight:
\begin{equation}
	x[n]
	\;\longrightarrow\;
	y[n]=\sum_{k=0}^{T-1}w[k]\,x[n-k]
	\;\longrightarrow\;
	Y=\sum_{n=0}^{N-1} y[n].
	\label{eq:thA:notation}
\end{equation}
Here $x[n]$ is the $n$-th complex sample of one readout shot, $w[k]$ is the FIR coefficient at tap index $k=0,\ldots,T-1$, $T$ is the tap count, and $y[n]$ is the filtered sample. $Y$ is the sum of the filtered samples over the $N$-sample window. Summing the raw samples and dividing by $N$ gives the boxcar average. $Y$ accumulates within one shot, while the training target $d[n]$ of Eq.~\ref{eq:inst:target} averages the same sample index across shots prepared in the same state. For the FIR with zero initial state, only the last $T-1$ input samples receive a truncated kernel. Every interior sample receives the same effective coefficient which is the kernel's DC gain. Apart from this finite-record edge, the filtered-and-summed statistic is a rescaled boxcar average statistic. We call this the kernel collapse. In the scalar-rescaling limit, a nonzero DC gain rotates and scales the accumulated I/Q clouds without improving their normalized separation. We call this the boxcar limit.

\begin{figure*}[!t]
  \centering
  \includegraphics[width=\textwidth]{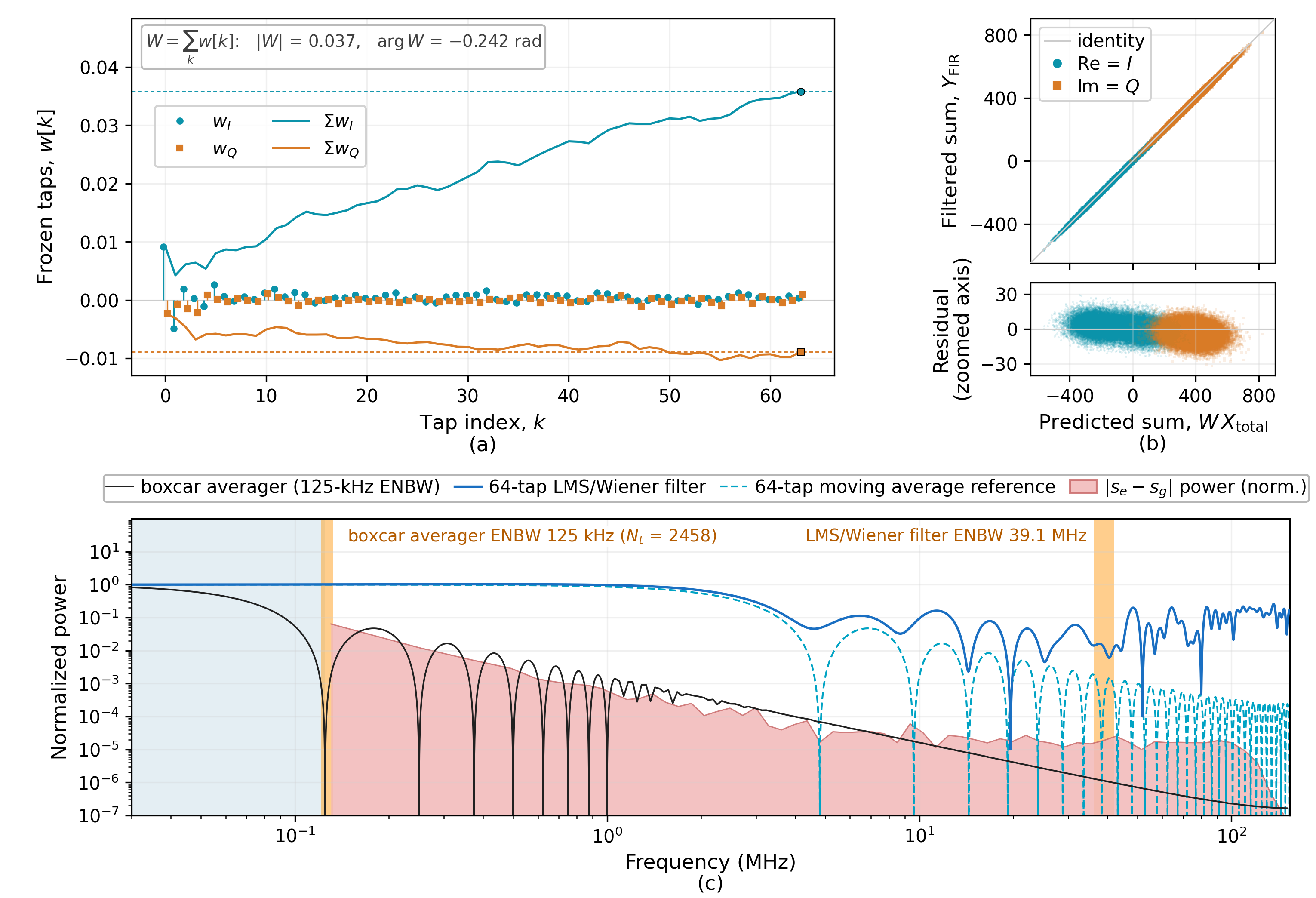}
  \caption
  	{
  	The kernel collapse on the deployed instrument [Q1 (transmon) at Readout Length (RL) 8~$\mu$s; 50,000 shots per class, half held out for testing].
  	(a) The frozen 64-tap coefficient vector, trained on-fabric with the qubit prepared in $|g\rangle$ and read back over DMA in Q27.16 fixed-point format. The running-sum overlay accumulates the taps to the collapse scalar $W$, with $|W| = 0.037$ and $\arg W = -0.242$~rad [exp.].
  	(b) The scalar approximation of Eq.~\ref{eq:thA:collapse}, tested on held-out shots from both classes. Each shot's filtered window sum $Y_{\rm FIR}$ is plotted against $W X_{\rm total}$, with $W$ from the taps in (a) and $X_{\rm total}$ the raw boxcar statistic of Eq.~\ref{eq:thA:collapse}; nothing was fitted. The points lie close to the identity line, so one complex scalar accounts for most of the rescaling and rotation. The lower strip shows the residual, with an RMS of 0.86\% of full scale. Its dependence on the predicted sum is consistent with a finite-record edge contribution [offline].
  	(c) The spectral budget. The black curve shows the boxcar averager's power response, drawn exactly below $\sim$1~MHz and as a peak-hold envelope above. Its two-sided equivalent noise bandwidth (ENBW) is 125~kHz. The interval between its first nulls at $\pm125$~kHz contains 93\% of the measured class-difference spectral power [offline]. The blue and dashed-cyan curves show the 64-tap LMS/Wiener and moving-average responses. The deployed FIR mainly removed high-frequency noise that the boxcar averager already strongly attenuates in the decision statistic.
  	}
  \label{fig:FIG-DT}
\end{figure*}

Fig.~\ref{fig:FIG-DT} anchors the result experimentally. This section derives the exact finite-record form, expresses its consequence in shot units, and determines the linear headroom available when the architecture realizes a time-dependent weight. The generalized matched filter sets the complex-output SNR ceiling. A separate $\cos^2\theta$ factor describes the possible loss when a sample-indexed readout trains on the direction of one state's average waveform instead of the state-difference direction.

\subsection{Trace SNR, Discrimination SNR, and Fidelity Uncertainty}
\label{sec:governing-snr}

We represent one readout shot of length $N$ by $\mathbf{x}\in\mathbb{C}^{N}$, a state-dependent waveform plus additive noise:
\begin{equation}
	\mathbf{x}=\mathbf{s}_q+\boldsymbol{\eta},\quad
	q=\text{state}\in\{g,e\},\quad
	\mathbf{R}_{nn}=\E[\boldsymbol{\eta}\boldsymbol{\eta}^{H}].
	\label{eq:thA:model}
\end{equation}
Equation~\ref{eq:thA:notation} describes one sample per clock, while Eq.~\ref{eq:thA:model} stacks the $N$ samples of one shot into $\mathbf{x}$. The binary label $q\in\{g,e\}$ denotes either the ground state $|g\rangle$ or the excited state $|e\rangle$, with $\mathbf{s}_q$ giving the corresponding mean readout waveform. $\E$ denotes expectation over shots, and $\mathbf{R}_{nn}$ is the covariance of the zero-mean noise $\boldsymbol{\eta}$, assumed common to both states in this model. A linear readout reduces the shot to the complex scalar $y=\mathbf{w}^{H}\mathbf{x}$, where $\mathbf{w}$ is the weight vector and $H$ denotes conjugate transpose. We define its complex-output discrimination SNR as~\cite{Kay1998,VanTrees2001}
\begin{equation}
	\mathrm{SNR}_{\rm disc}(\mathbf{w})=
	\frac{\left|\mathbf{w}^{H}(\mathbf{s}_e-\mathbf{s}_g)\right|^2}
		 {\mathbf{w}^{H}\mathbf{R}_{nn}\mathbf{w}}.
	\label{eq:thA:snr-disc}
\end{equation}

We distinguish trace SNR, discrimination SNR, and the standard error of the fidelity estimate. $\mathrm{SNR}_{\rm trace}$ describes the signal relative to noise in a single trace. $\mathrm{SNR}_{\rm disc}$ measures the squared separation of the class means relative to the noise variance of the pre-threshold score. $\mathrm{SE}(\widehat{F}_M)$ gives the uncertainty in the fidelity estimated from $M$ labeled shots per class using a fixed classifier. At the measured operating points, the deployed filter improved $\mathrm{SNR}_{\rm trace}$ without a systematic improvement in $\mathrm{SNR}_{\rm disc}$ or a reduction in the validation shots needed for a given fidelity standard error.
The full-window boxcar averager has $\mathbf{w}=\mathbf{1}$, where $\mathbf{1}$ is the all-ones vector. For a positive-definite noise covariance, the generalized matched filter and its maximum complex-output SNR are
\begin{equation}
	\begin{aligned}
	\mathbf{w}_{\rm disc}
	&=\mathbf{R}_{nn}^{-1}(\mathbf{s}_e-\mathbf{s}_g),\\
	\mathrm{SNR}_{\rm max}
	&=(\mathbf{s}_e-\mathbf{s}_g)^H\mathbf{R}_{nn}^{-1}
	(\mathbf{s}_e-\mathbf{s}_g).
	\end{aligned}
	\label{eq:thA:mf}
\end{equation}
No linear weight exceeds this value for the SNR defined in Eq.~\ref{eq:thA:snr-disc}~\cite{Kay1998,VanTrees2001}.

An independent, noise-only reference signal correlated with the readout chain noise could provide another route. However, that escape as adaptive noise cancellation (ANC) requires an additional reference input that was not built into the setup. It would also likely provide negligible advantage over the deployed filter since the amplifier-added noise that dominates the averager passband has no signal-free copy to cancel. Thus, we trained toward an ensemble-average signal target as this better fit the hardware constraints (Sec.~\ref{sec:instrument}).

\subsection{The Kernel Collapse}
\label{sec:governing-law}

For the filter-and-sum composition of Eq.~\ref{eq:thA:notation}, with any $T$ coefficients $w[k]$, exchanging the two sums gives
\begin{equation}
	\begin{aligned}
	Y
	&=\sum_n\sum_{k=0}^{T-1}w[k]\,x[n-k]\\
	&\approx
	\left(\sum_{k=0}^{T-1}w[k]\right)\sum_n x[n]
	=WX_{\rm total}.
	\end{aligned}
	\label{eq:thA:collapse}
\end{equation}
Here, $W=\sum_{k=0}^{T-1}w[k]$ is the filter's DC gain and $X_{\rm total}=\sum_n x[n]$ is the raw boxcar statistic. This is the classical input/output identity for the mean of a filtered process~\cite[Sec.~5.2]{GrayDavisson2004}.
The coefficients' origin does not matter: $w[k]$ may be the on-fabric-trained LMS/Wiener taps, a host-computed design, or the moving average reference, and the collapse follows from the composition alone.
The contribution here is to identify it as the governing decision rule of the deployed readout path, derive its exact finite-record form, and verify it on measured shots.

Substituting the scaling of Eq.~\ref{eq:thA:collapse} into Eq.~\ref{eq:thA:snr-disc} gives
\begin{equation}
	\mathrm{SNR}_{\rm disc,FIR}
	=
	\frac{|W|^2\left|\mathbf{1}^{H}(\mathbf{s}_e-\mathbf{s}_g)\right|^2}
     	 {|W|^2\mathbf{1}^{H}\mathbf{R}_{nn}\mathbf{1}}
	=
	\mathrm{SNR}_{\rm disc,box}.
	\label{eq:thA:scale-invariance}
\end{equation}
The factor $|W|^2$ cancels because the same complex scaling multiplies the state separation and the noise variance. Trace cleanup therefore does not imply a better decision statistic.

The following form is exact for a causal, frozen FIR filter with zero initial state before fixed-point rounding or truncation. Let $C_w$ be the convolution matrix for the $T$ frozen taps over the $N$-sample record. $\mathbf{a}$ denotes the accumulation-weight vector with one entry for each filtered sample. For uniform full-window summation every entry equals one, so $\mathbf{a}=\mathbf{1}$. Writing $Y_{\rm FIR}$ for the $Y$ of Eq.~\ref{eq:thA:notation},
\begin{equation}
	Y_{\rm FIR}=\mathbf{a}^{H}C_w\mathbf{x}
	=\left(\mathbf{w}^{\rm eff}_{\rm FIR}\right)^{H}\mathbf{x}.
	\label{eq:thA:exact}
\end{equation}
The effective decision-weight vector is $\mathbf{w}^{\rm eff}_{\rm FIR}=C_w^{H}\mathbf{a}$. Under these boundary conditions, the identity holds for any waveform within the record.
For this uniform window, the row vector $\mathbf{a}^{H}C_w$ applies the same coefficient $W$ to every input sample except possibly the final $T-1=63$ samples. Those samples have filter responses that extend beyond the end of the accumulation window. At RL~$=8~\mu$s, the measured fraction of entries differing from $W$ was $0.0256$, matching $(T-1)/N=63/2458$. The pre-registered estimate $T/N=0.0260$ counts one extra sample. For fixed $T$, the fraction of potentially affected entries decreases as $1/N$. The bit-true twin, a fixed-point software model of the hardware, checked the additional arithmetic error at the hardware truncation floor (Sec.~\ref{sec:instrument-verify}).

Fig.~\ref{fig:FIG-DT}(b) tests the scalar approximation shot by shot. Fig.~\ref{fig:FIG-DT}(c) shows the spectral overlap at RL~$=8~\mu$s. Frequencies are baseband offsets from the readout carrier after digital down-conversion to the 307.2-MS/s decimated stream.
The boxcar averager has a two-sided ENBW of 125~kHz. The 64-tap moving-average reference and frozen LMS/Wiener filter had ENBWs of 4.8 and 39.1~MHz, respectively. Meanwhile, $93$\% of the measured class-difference spectral power lay within $\pm125$~kHz of DC, including $78$\% in the DC bin. The FIR removed high-frequency trace noise that the boxcar averager already strongly attenuates in the decision statistic.
A unit-DC-gain $T$-tap FIR has a two-sided ENBW of at least $f_s/T$, where $f_s$ is the sample rate. Equal taps attain this bound. Matching the accumulator's ENBW therefore requires at least $N=2458$ taps. At the same parallel update architecture, this requires approximately 38 times as many tap updates as the deployed 64-tap filter.

Same-shot offline replay of the frozen 64-tap LMS/Wiener filter shows no systematic gain over the boxcar averager at any live readout length ($|W|=0.037$--$0.065$). The causal 64-tap kernel tested offline, obtained by projecting the whitened per-sample matched filter onto the convolutional basis, gave $+0.00$\% averaged over all readout lengths and $+0.24$\% over the 2--8-$\mu$s band of the sweep.
The 64-tap moving average reference still removes $16.2$--$16.3$~dB of trace noise (Sec.~\ref{sec:instrument-denoise-char}) while its RL8 fidelity change is consistent with zero ($-0.16$\%, 95\% confidence interval (CI) $[-0.48,+0.13]$). No tested coefficient set converts the finite edge into a reusable gain at these operating points.

Escaping the collapse requires changing the composition rather than improving convergence within it. A shift-invariant FIR can implement a matched filter if one designated output sample is used instead of accumulating the complete
output record. A direct sample-indexed accumulator can realize
\begin{equation} 
	Y_{\rm weighted}=\sum_n w^*[n]x[n]=\mathbf{w}^{H}\mathbf{x}.
	\label{eq:thA:wgt}
\end{equation}

\subsection{The Equivalent-Shot Currency}
\label{sec:governing-meq}

The operational consequence is most directly expressed in shots. Define
\begin{equation}
	M_{{\rm eq},m}=
	\frac{\mathrm{SNR}_{{\rm disc},m}}
     	 {\mathrm{SNR}_{\rm disc,box}}.
	\label{eq:thA:meq}
\end{equation}
Here $m$ indexes the detector, and $M_{{\rm eq},m}$ is its dimensionless discrimination-SNR ratio relative to the boxcar. The empirical SNRs used in this ratio are defined from held-out real scores in Eq.~\ref{eq:thA:deflection}. For independent repetitions at equal accumulated deflection, the required number of readouts scales as $1/M_{{\rm eq},m}$ relative to the boxcar. This is not the number of shots needed to estimate a fixed classifier's fidelity. At RL~$=8~\mu$s, the frozen-filter detector gave $0.9945$ (95\% CI $[0.9931,0.9960]$). The values at RL~$=4$, 6, and 13~$\mu$s were $1.0192$, $0.9910$, and $0.9934$, respectively, never more than two percent from unity.
The sign of their departure from unity changes with readout length, consistent with a finite-window edge rather than a systematic gain. By contrast, the per-sample matched filter gives $M_{\rm eq}=1.20$ at the shortest window and $0.977$ at the longest, where the deployed calibration budget no longer resolves an advantage (Sec.~\ref{sec:map}).

\subsection{Escaping the Collapse: The Matched-Filter Ceiling}
\label{sec:governing-ceiling}

The kernel-collapse result of Sec.~\ref{sec:governing-law} bounds the deployed FIR-plus-boxcar-averager composition. When a linear readout instead realizes a general sample-indexed weight, the boxcar averager reaches the matched-filter ceiling exactly when $(\mathbf{s}_e-\mathbf{s}_g)$ is proportional to $\mathbf{R}_{nn}\mathbf{1}$. The generalized matched filter strictly improves on it if and only if that proportionality fails. This condition identifies two physical sources of linear headroom. A third subsection treats the penalty that survives even when the headroom is realized.

At the transmon operating point, a Gambetta-convention calculation places the full matched-filter ceiling approximately $+1.6$\% above the boxcar averager, whereas the associated measured comparison is $+0.12$\%. The $+1.6$\% is a calculated ceiling, not a measured gain. These statements are operating-point scoped: a spur-rich chain or strongly nonstationary waveform changes both the calculated ceiling and the measured comparison.

\subsubsection{Colored Noise}
\label{sec:governing-colored-noise}

If the noise covariance $\mathbf{R}_{nn}$ is not proportional to the identity $\mathbf{I}$ in the information-bearing band, whitening can reweight noisy directions. This source of headroom is nearly absent on the transmon. Its in-band noise is approximately white, while measurable color lies at 20--150~MHz and is spectrally separated from the readout band. The computed whitening bound is approximately $+0.6$\%, conditioned on drift residing in the removable per-class mean. 
A 256-bin whitened matched filter evaluated on the same shots [offline] reports a $+0.4$ to $+2.4$\% contrast margin over the per-sample matched filter, equivalent to $+0.2$ to $+1.2$\% on the mean-assignment scale. On the full-rate QICK acquisition of the ancilla qubit (Q2), the pre-registered whitening test gives $+0.23$\% (95\% CI $[+0.12,+0.35]$) against a $+1.0$\% pre-registered threshold. A planted $+1.8$\% synthetic effect survives the same pipeline, so the null is not caused by an insensitive analysis.

\subsubsection{Time-Varying State Difference}
\label{sec:governing-time-varying}

If $(\mathbf{s}_e-\mathbf{s}_g)[n]$ varies over the window because of ring-up (the transient buildup of the resonator field) or $T_1$ decay, sample-indexed weights can outperform a flat profile. This headroom is present but concentrated at short readout lengths. On the live acquisition, the per-sample matched filter gains $+3.00/+1.90/+2.62$\% at RL~$=4/6/8~\mu$s and ties at RL~$=13~\mu$s ($+0.02$\%, CI including zero). The earlier sweep gives $+3.26$\% at RL~$=4~\mu$s (95\% CI $[+2.48,+4.02]$\%), a CI including zero at RL~$=10~\mu$s, and a negative result at RL~$=13~\mu$s (CI $[-1.43,-0.40]$\%). Those earlier values use mean-assignment fidelity; the contrast-scale differences are twice as large.

Against the tuned boxcar averager, the residual advantage at RL~$=8~\mu$s is $+0.33$\%. Most of the gain over full-window accumulation therefore comes from the integrator window placement. 
The gain depended on which samples the window retained. The tuned boxcar averager removed early samples before the resonator response developed. Selecting only the relaxation tail gave a negative fidelity change at RL~$=6$ and 8~$\mu$s. The RL13 tie reflected the deployed template-training budget rather than a fundamental ceiling (Sec.~\ref{sec:map}).

\subsubsection{Training-Objective Mismatch and the $\cos^2\theta$ Penalty}
\label{sec:governing-cos2}

A sample-indexed readout can still use a training objective that does not maximize state separation. A minimum-mean-square-error (MMSE) filter trained on an average per-state waveform minimizes reconstruction error, which differs from the discrimination objective in Eq.~\ref{eq:thA:mf}. To describe a possible direction loss, consider a readout weight $\mathbf{w}_{\rm avg}$ proportional to $\mathbf{R}_{nn}^{-1}\mathbf{s}_{\rm avg}$, where $\mathbf{s}_{\rm avg}$ is the mean waveform for that prepared state and the training target $d[n]$ in Eq.~\ref{eq:inst:target}. Define $\theta$ by
\[
\cos\theta=
\frac{\left|\mathbf{s}_{\rm avg}^{H}\mathbf{R}_{nn}^{-1}(\mathbf{s}_e-\mathbf{s}_g)\right|}
{\sqrt{\left(\mathbf{s}_{\rm avg}^{H}\mathbf{R}_{nn}^{-1}\mathbf{s}_{\rm avg}\right)
\left((\mathbf{s}_e-\mathbf{s}_g)^H\mathbf{R}_{nn}^{-1}(\mathbf{s}_e-\mathbf{s}_g)\right)}}.
\]
Here $\theta$ is the phase-insensitive angle between the two nonzero templates in the $\mathbf{R}_{nn}^{-1}$-weighted inner product. Substitution gives
\begin{equation}
	\frac{\mathrm{SNR}_{\rm disc}(\mathbf{w}_{\rm avg})}
     	 {\mathrm{SNR}_{\rm disc}(\mathbf{w}_{\rm disc})}
	=\cos^2\theta.
	\label{eq:thA:cos2}
\end{equation}
Better convergence does not remove the loss from a fixed, mismatched template. Gray and Davisson Thm.~4.9 gives the orthogonality principle for linear MMSE estimation~\cite{GrayDavisson2004}. Here, the SNR ratio follows directly from the two whitened template directions. In the two-dimensional dispersive-drive model, the computed penalty is approximately $0.96$~dB at $\chi/\kappa=1.0$, with $\chi$ the dispersive shift, and spans 1.0--12~dB over $\chi/\kappa\in[0.1,1.0]$~\cite{Gambetta2008,Blais2021}. Here, $\kappa/2\pi$ is the readout-resonator linewidth, with field-amplitude time constant $2/\kappa$. It binds on the per-sample escape path, where direction matters. On the collapsed FIR path, scalar invariance has already made direction irrelevant.

\subsection{Empirical Closure}
\label{sec:governing-closure}

For each fixed linear detector $m$, the effective weight determines the real pre-threshold score. Before derotation, this weight is $C_w^H\mathbf{a}$ for the FIR-plus-accumulator composition of Eq.~\ref{eq:thA:exact}, the design weight for the sample-indexed detector of Eq.~\ref{eq:thA:wgt}, and the all-ones vector for the boxcar. The training-fixed derotation is included in the effective detector weight, giving $r_m=\operatorname{Re}\{(\mathbf{w}_m^{\rm eff})^H\mathbf{x}\}$.
The state separation is $\delta_m=\lvert\E[r_m\mid e]-\E[r_m\mid g]\rvert$, set by $\chi$ against $\kappa$~\cite{Krantz2019}, and the pooled within-state spread is $\sigma_m$, with $\sigma_m^{2}=\tfrac{1}{2}(\sigma_{m,g}^{2}+\sigma_{m,e}^{2})$ and $\sigma_{m,q}^{2}=\Var(r_m\mid q)$. The reported metrics are
\begin{equation}
	\left(\frac{\delta}{\sigma}\right)_m=
	\frac{\delta_m}{\sigma_m},
	\qquad
	\mathrm{SNR}_{{\rm disc},m}=
	\left(\frac{\delta}{\sigma}\right)_m^2.
	\label{eq:thA:deflection}
\end{equation}
Equation~\ref{eq:thA:snr-disc} defines the model SNR of the complex output. Equation~\ref{eq:thA:deflection} defines the empirical real-score SNR used for the reported equivalent-shot ratios. Numerators and denominators use the same scoring convention. Per-sample plots evaluate $\delta$ and $\sigma$ at each sample $n$ and write $(\delta/\sigma)[n]$.

The maximum measured in-band I/Q impropriety was $0.11$. We therefore used the full $2\times2$ I/Q cross-spectral density matrix in the frequency-domain analysis. Held-out time-domain projection provided the primary quantitative check.
The analysis used an additive-noise model at the saved digital reference plane and assumed within-session stationarity. It retained I/Q correlations, used the measured finite window, and checked the frozen linear model against the fixed-point implementation.

\subsection{The Conjecture's Verdict}
\label{sec:governing-verdict}

The original premise is confirmed: the adaptive instrument removes noise from each trace. However, the inference built on it fails, and the measured record lands where Secs.~\ref{sec:governing-law}--\ref{sec:governing-ceiling} predict with no systematic gain from the LMS/Wiener filter. The shot-efficiency claim falls with the fidelity claim. Under the collapse, each filtered statistic is a fixed complex rescaling of its raw counterpart, so the leading shot requirement of a fixed model-free estimator is unchanged. All sixteen fitted log--log slope CIs contain $-1/2$, consistent with the same $M^{-1/2}$ scaling. The fewer-shot objective survives instead in model-based calibration, which estimates decision templates from tens of labeled shots. 
The remaining claim that adaptation improves calibration visibility, threshold stability, or drift tracking, cannot be settled with the current architecture and is saved for future work.

\section{The On-Fabric Adaptive Filter as a Measurement Instrument}
\label{sec:instrument}

The adaptive filter was designed and implemented on fabric as a custom Quantum Instrumentation Control Kit (QICK) readout module~\cite{Stefanazzi2022}. This section presents its model and implementation footprint, verifies its training, convergence, and coefficient readback during qubit operation, and quantifies the removable-noise budget when the objective shifts from discrimination to estimation.

For labeled acquisitions, we used no $\pi$ pulse for $|g\rangle$ and a calibrated $\pi$ pulse to prepare $|e\rangle$ before readout. We trained the filter on measured readout traces for a prepared state and read back its frozen coefficients from the fabric. We computed the collapse scalar $W$ of Eq.~\ref{eq:thA:collapse} from those coefficients. A bit-true software model checked the fixed-point datapath and its truncation error. The integer update has an all-zero failure mode that is absent from the unquantized LMS model. Replaying the read-back coefficients on the same shots showed no resolved fidelity improvement over the boxcar averager. 
Here [exp.] denotes hardware measurements, [bit-true twin] denotes fixed-point model checks, and [offline] denotes analysis of measured shots after acquisition.

\subsection{The Instrument Model}
\label{sec:instrument-model}

\begin{figure}[t]
	\centering
	\includegraphics[width=\columnwidth]{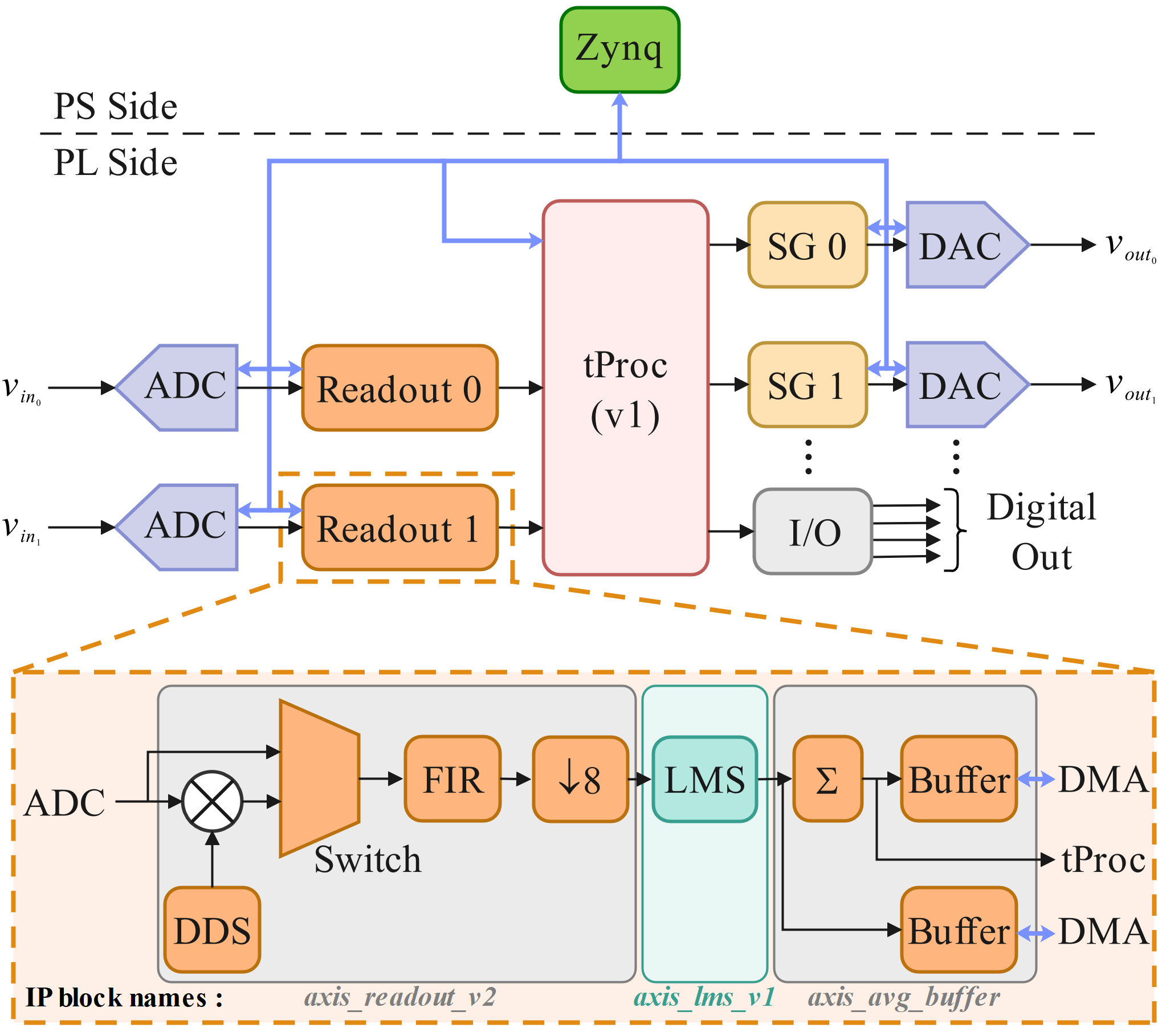}
	\caption{
	The instrument in the QICK readout chain. The Zynq Processing System (PS) accesses the readout and control blocks in the Programmable Logic (PL). These blocks connect the converters (ADC, DAC), two readout channels, timed processor (tProc), signal generators, and digital I/O. The expanded readout-1 path shows DDS downconversion, FIR filtering, and decimation by eight (\texttt{axis\_readout\_v2}) feeding the LMS filter (\texttt{axis\_lms\_v1}, teal; Fig.~\ref{fig:FIG-D2}). Its output enters the accumulate-and-buffer stage (\texttt{axis\_avg\_buffer}), and the windowed sum feeds the tProc. The buffers transfer data to the host by Direct Memory Access (DMA). Blue arrows mark PS-side access. Panels redrawn from~\cite{Stefanazzi2022} with the adaptive filter inserted into the digital readout chain.
	}
	\label{fig:FIG-D1}
\end{figure}

\begin{figure}[H]
	\centering
	\setlength{\abovecaptionskip}{4pt}
	\includegraphics[width=\columnwidth]{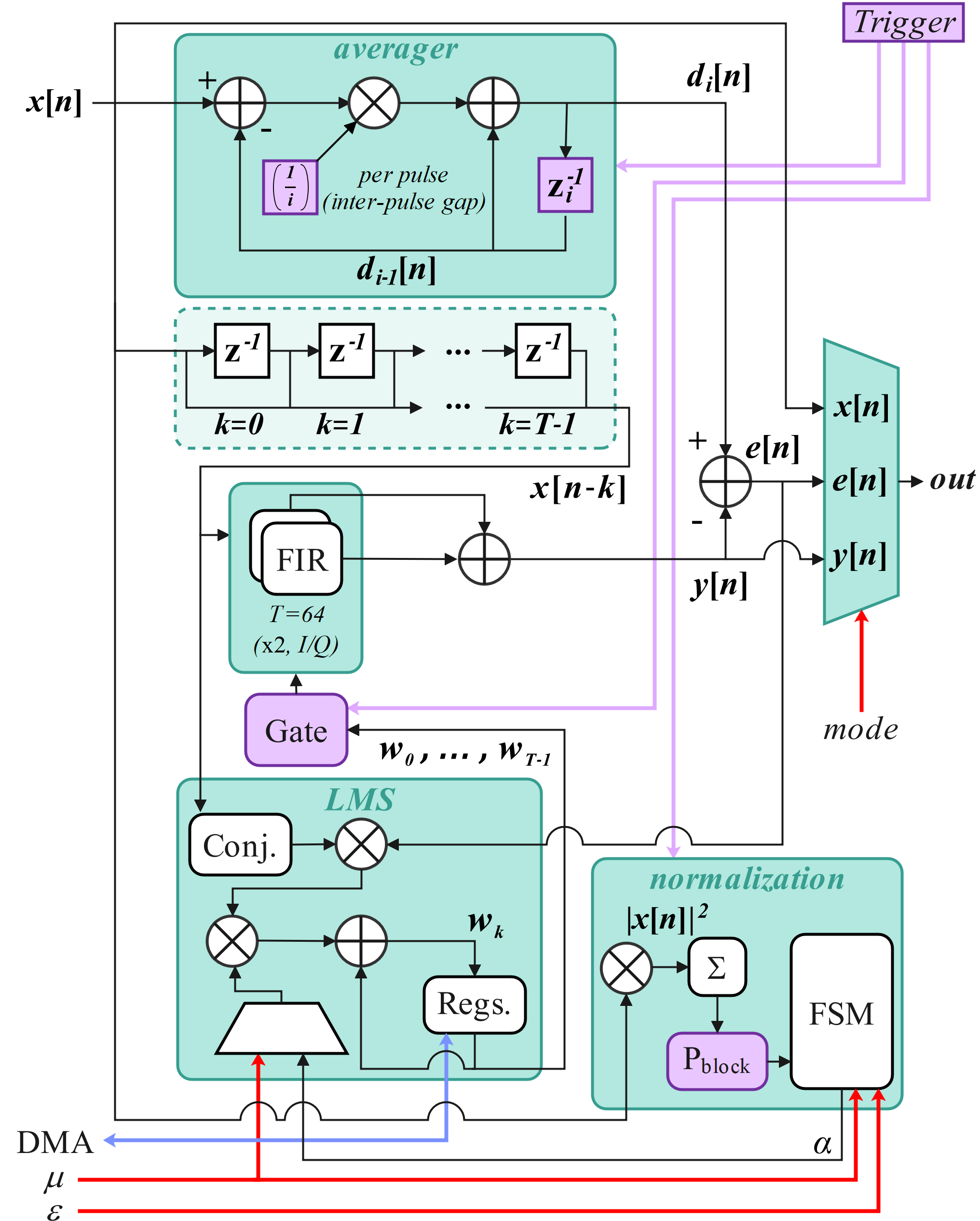}
	\caption{
	Adaptive filter (\texttt{axis\_lms\_v1}) firmware block diagram. The averager forms the reference $d_i[n]$ across pulses ($z_i^{-1}$: a one-pulse delay at fixed sample position $n$). The FIR filter and the LMS blocks share one tapped delay line. The LMS block conjugates the regressor, scales the gradient, and integrates it into the weight registers. A pulse-boundary gate releases the updated weights to the FIR, closing the adaptation loop. A reciprocal Finite-State Machine (FSM) forms the per-pulse step $\alpha$ in the normalization block (Eq.~\ref{eq:inst:alpha}). Black lines indicate AXI4-Stream (AXIS) datapaths, red lines denote run-time user input via AXI4-Lite (AXI), blue lines are DMA PS-side access, and purple denotes per-pulse hardware operations initiated by a tProc trigger.
	}
	\label{fig:FIG-D2}
\end{figure}

Figs.~\ref{fig:FIG-D1} and~\ref{fig:FIG-D2} show the hardware as high-level block designs. The filter exposes mode and NLMS\_BASE\_SHIFT controls, a per-pulse Mean-Squared-Error (MSE) status register, and a DMA interface for coefficient readback described in Sec.~\ref{sec:instrument-verify}. The datapath is a $T=64$-tap complex FIR over the I/Q sample stream,
\begin{equation}
	y[n] = \sum_{k=0}^{T-1} w_k[n]\, x[n-k],
	\label{eq:inst:fir}
\end{equation}
trained against the ensemble-average target $d[n]$ that the on-fabric averager (Fig.~\ref{fig:FIG-D2}) builds by the Welford update
\begin{equation}
	d_i[n] = d_{i-1}[n] + \frac{x[n] - d_{i-1}[n]}{i},
	\label{eq:inst:target}
\end{equation}
with the error driving adaptation:
\begin{equation}
	e[n] = d[n] - y[n].
	\label{eq:inst:error}
\end{equation}
In Eq.~\ref{eq:inst:fir}, $w_k[n]$ denotes the direct complex coefficient stored by the firmware. Eq.~\ref{eq:inst:target} and Fig.~\ref{fig:FIG-D2} uses $i$ to represent the readout pulse count. After convergence and freezing in Mode-1, these coefficients become the fixed $w[k]$ of Eqs.~\ref{eq:thA:notation} and~\ref{eq:thA:collapse}.

The weight update, implemented by the LMS branch of Fig.~\ref{fig:FIG-D2}, is the block-normalized LMS step
\begin{equation}
	\mathbf{w}[n+1] = \mathbf{w}[n] + \alpha_{\rm eff}\, e[n]\,
	\mathbf{x}_n^{*},
	\qquad
	\alpha_{\rm eff} = \frac{\alpha}{2^{S_g}},
	\label{eq:inst:update}
\end{equation}
where $\mathbf{x}_n = [x[n], \ldots, x[n-T+1]]^T$ is the tap vector, $\mathbf{w}[n]$ collects the $T$ weights of Eq.~\ref{eq:inst:fir}, and $S_g$ is the barrel-shift gradient prescaler. The normalization branch forms the step once per pulse from the window's accumulated power,
\begin{equation}
	\alpha = \frac{\mu}{\varepsilon' + P_{\rm block}},
	\qquad
	P_{\rm block} = \sum_{n=0}^{N-1} \bigl| x[n] \bigr|^2,
	\label{eq:inst:alpha}
\end{equation}
with $\mu$ the programmed learning rate and $\varepsilon'$ the regularization constant. Mean convergence requires $0 < \alpha_{\rm eff} < 2/\lambda_{\max}$, where $\lambda_{\max}$ is the largest eigenvalue of the regressor covariance matrix. The implemented filter commits coefficients in between pulses.
The firmware uses the direct-coefficient, conjugate-regressor form of Eqs.~\ref{eq:inst:fir} and~\ref{eq:inst:update}.
The decision-weight vector is the element-wise conjugate of the stored coefficient vector only when the same linear map is written in the abstract $\mathbf{w}^{H}\mathbf{x}$ convention of Sec.~\ref{sec:governing-snr}. A mode register routes the output:
\begin{equation}
	\text{Output} =
	\begin{cases}
	y[n],\ w~\text{adapting} & \text{Mode-0 (LMS)}\\[2pt]
	y[n],\ w~\text{frozen} & \text{Mode-1 (Wiener)}\\[2pt]
	x[n] & \text{Mode-2 (bypass)}\\[2pt]
	e[n] & \text{Mode-3 (error)},
	\end{cases}
	\label{eq:inst:modes}
\end{equation}
The bypass is a registered passthrough for the unfiltered signal.

For the deployed frozen filter (Mode-1, Wiener filter), the accumulated output is the boxcar sum multiplied by $W$, apart from the finite-window edge correction and fixed-point error (Eq.~\ref{eq:thA:collapse}). The state separation and noise spread both scale by $|W|$, so the factor cancels from the deflection as shown in Eq.~\ref{eq:thA:scale-invariance}. To reiterate, this means the qubit state I/Q cloud's (such as Fig.~\ref{fig:FIG-DT}(b)) center of mass move closer together by $|W|$ while the blobs shrink by $|W|$, and the ratio of $(\delta/\sigma)$ is unchanged. The filter reduced single-shot trace noise without improvement to single-shot fidelity at these operating points because the removed noise power is out-of-band to the accumulated measured record (Sec.~\ref{sec:governing-law}).

Additionally, it is important to distinguish single-shot readout fidelity in two conventions: contrast and mean-assignment. Early circuit-QED work typically report contrast~\cite{Gambetta2007,Walter2017,Blais2021}, and engineering/error-correction literature typically reports mean-assignment~\cite{Magesan2015,Martinez2020,DiGuglielmo2025}. This paper focuses on contrast readout fidelity because it presents a more visually digestible reading for near-chance operating points, which is what this paper is centered on comparing. However, the two report identical data and are related by $F_{\rm ma} = (1 + F_c)/2$. The contrast convention is
\begin{equation}
	\begin{gathered}
	F_c = 1 - P(e|g) - P(g|e),
	\end{gathered}
	\label{eq:inst:fidelity-c}
\end{equation}
which sits at zero for a detector at chance. The mean-assignment convention is
\begin{equation}
	\begin{gathered}
	F_{\rm ma} = 1 - \tfrac{1}{2}\bigl[ P(e|g) + P(g|e) \bigr].
	\end{gathered}
	\label{eq:inst:fidelity-ma}
\end{equation}
Here $P(e|g)$ is the probability of assigning $e$ after preparing $g$, and $P(g|e)$ is defined analogously. The mean-assignment convention gives equal weight to the two prepared classes. A state-independent classifier has $F_c = 0$ and $F_{\rm ma} = 50\%$. Fidelity changes are absolute differences expressed in \%, while shot-count savings are relative reductions.

\subsection{Implementation Footprint}
\label{sec:instrument-footprint}

Table~\ref{tab:footprint} summarizes the IP footprint. The parallel 64-tap complex LMS update uses 512 DSP48E2 slices, with eight per tap for the gradient, step scaling, and accumulation. Most of the remaining DSP use comes from the two vendor FIR cores, with fewer than a dozen used for normalization, averaging, and error telemetry. At 1,558 slices per instance, three instances would exceed the ZCU216 FPGA's 4,272-slice capacity before adding the remaining QICK logic. The stream accepts and produces one complex sample per clock, without backpressure.

\begin{figure}[H]
	\centering
	\includegraphics[width=0.72\columnwidth]{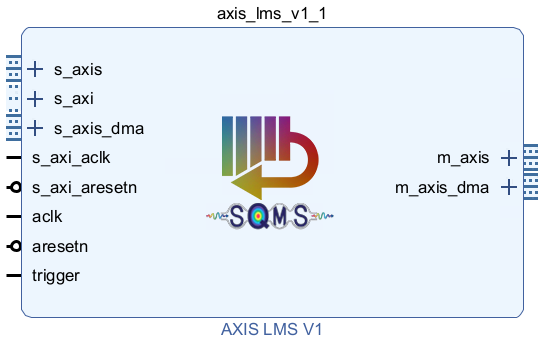}
	\caption{
	Packaged adaptive-filter IP in the Vivado block-design integrator. AXI4-Stream ports carry readout samples input/output and coefficient transfers with an external DMA. AXI4-Lite provides user-defined variables via register access during experiments. The tProc trigger coordinates pulse-boundary operations. Stream and configuration interfaces have separate clock/reset pairs.
	}
	\label{fig:FIG-D3}
\end{figure}

The custom LMS update takes 18 clocks and runs in parallel with filtering. The vendor FIR cores contribute 79 of the filtered path's 86 clocks, so they dominate its 280-ns latency at 307.2~MHz, not its sample throughput. Coefficient reload, reciprocal calculation, and MSE processing occur between pulses. Optimizing the FIR implementation is a possible route to lower resource use or latency, but this was not evaluated here. Fig.~\ref{fig:FIG-D3} shows the packaged interface.

\begin{table}[t]
	\caption{Resource Utilization and Latency of the Adaptive Filter IP}
	\label{tab:footprint}
	\centering
	\footnotesize
	\setlength{\tabcolsep}{3pt}
	\begin{tabular}{@{}lrrrr@{}}
	\toprule
	\multicolumn{5}{@{}l}{\emph{Utilization (post-synthesis, in-context; XCZU49DR, ZCU216 RFSoC)}}\\
	Resource & Used & Available & \multicolumn{2}{r}{Util.\ (\%)} \\
	\midrule
	Look-Up Tables (LUTs) & 52{,}244 & 425{,}280 & \multicolumn{2}{r}{12.3} \\
	Flip-Flops (FFs) & 47{,}664 & 850{,}560 & \multicolumn{2}{r}{5.6} \\
	DSP48E2 & 1{,}558 & 4{,}272 & \multicolumn{2}{r}{36.5} \\
	Block-RAM (BRAM) tiles & 16.5 & 1{,}080 & \multicolumn{2}{r}{1.5} \\
	\midrule
	\multicolumn{5}{@{}l}{\emph{Latency (IP-internal; data clock 307.2~MHz)}}\\
	Path & Clocks & \multicolumn{3}{r}{Time} \\
	\midrule
	Bypass ($x[n]$, Mode-2) & 2 & \multicolumn{3}{r}{6.5~ns} \\
	Filtered ($y[n]$, Mode-0/1) & 86 & \multicolumn{3}{r}{280~ns} \\
	Error ($e[n]$, Mode-3) & 90 & \multicolumn{3}{r}{293~ns} \\
	\midrule
	\multicolumn{5}{@{}l}{\emph{By subsystem (RTL stages; streaming / per-pulse)}}\\
	& \multicolumn{2}{c}{Streaming} & \multicolumn{2}{c}{Per-pulse} \\
	\cmidrule(lr){2-3}\cmidrule(lr){4-5} Subsystem & Clocks & Time & Clocks & Time \\
	\midrule
	LMS weight update (64 taps) & 18 & 58.6~ns & -- & -- \\
	NLMS power/reciprocal & 4 & 13.0~ns & 6 & 19.5~ns \\
	Welford reference averager & 12 & 39.1~ns & -- & -- \\
	Error calculation & 4 & 13.0~ns & 4 (MSE) & 13.0~ns \\
	Vendor FIR cores (I/Q) & 79 & 257~ns & 128 (reload) & 417~ns \\
	Splitter + mode mux & 2 & 6.5~ns & -- & -- \\
	Output recombiner & 5 & 16.3~ns & -- & -- \\
	\bottomrule
	\end{tabular}
\end{table}

\subsection{Experimental Verification on the Qubit}
\label{sec:instrument-verify}

Quantization creates an update dead zone absent from the unquantized LMS model. Q notation is used to define integer bits and fractional bits. The Q27.16 coefficient registers have 27 integer bits, and 16 fractional bits corresponding to an LSB of $2^{-16}$ in coefficient units. Each real and imaginary component of the Normalized-Least-Mean-Squares (NLMS) update in Eq.~\ref{eq:inst:update} is truncated toward zero to this resolution:
\begin{equation}
	w_k[n+1] = w_k[n] +
	\operatorname{trunc}\bigl[\alpha_{\rm eff}\, e[n]\,
	x^{*}[n-k]\bigr]_{\rm LSB}.
	\label{eq:inst:floor}
\end{equation}

The consequence is a silent failure. An update smaller than one LSB truncates to zero, so sub-LSB increments never accumulate. Because the coefficients initialize to zero, a filter driven below this amplitude floor trains exactly all-zero coefficients while every live status indicator reads healthy. The programmed shift total $S_b + S_g$ sets the floor, and every qubit operating point in this record is well above it. Training loss does not expose the failure either because a large error drop can accompany oscillation as easily as convergence. The per-pulse MSE register flags divergence by reading the error growth directly, and the DMA coefficient read-back is the only ground truth for what the instrument will actually apply. For fixed-point adaptive filtering, coefficient telemetry, not the loss curve, is the health metric.

\begin{figure}[t]
  \centering
  \includegraphics[width=\columnwidth]{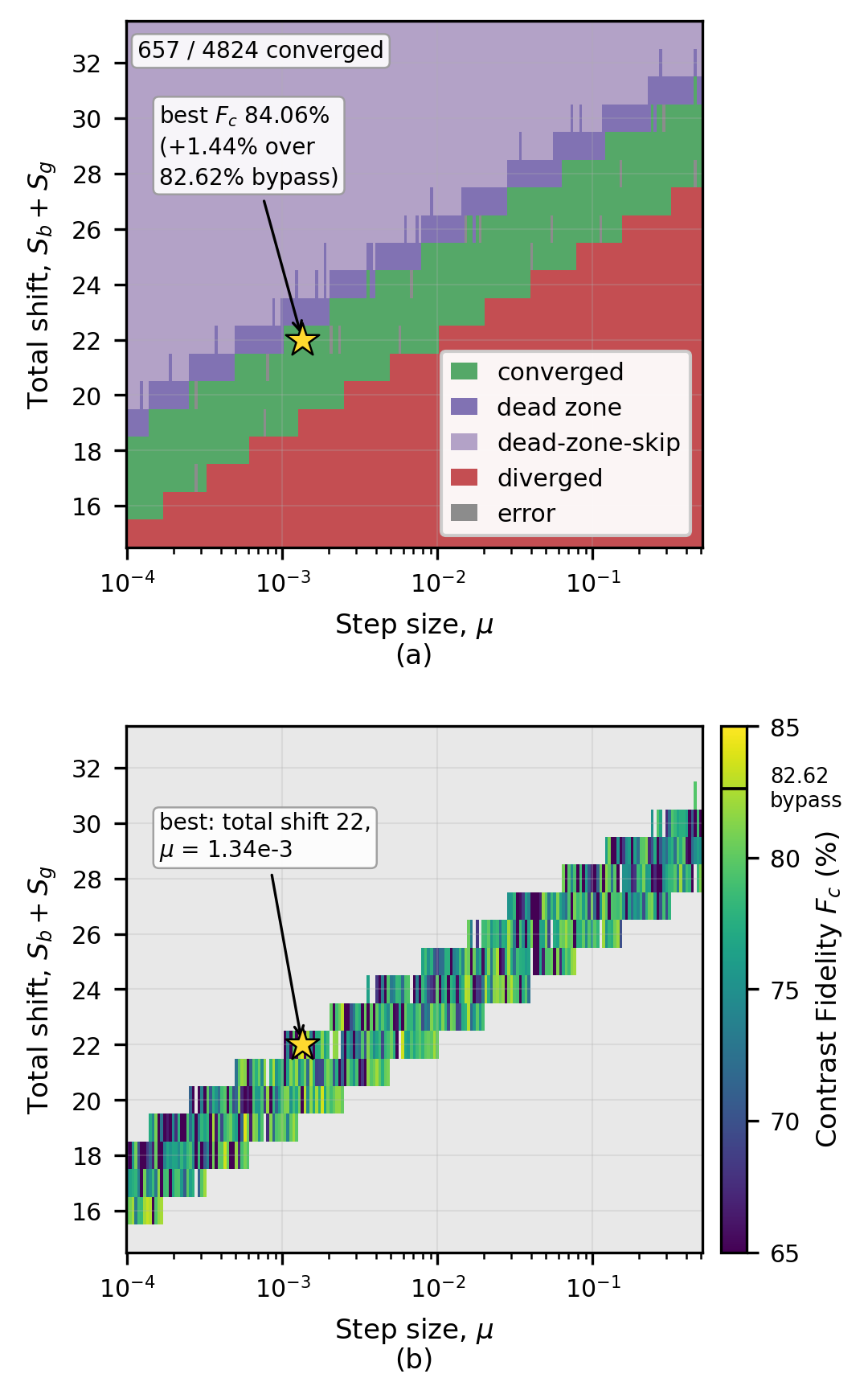}
  \caption{
  The convergence regime of the transmon (Q1) training sweep [exp.].
  (a) Regime map over step size $\mu$ and total shift $S_b + S_g$, each configuration classed as converged, dead-zone (the amplitude-floor regime of Eq.~\ref{eq:inst:floor}), dead-zone-skip, diverged, or error. The 1514 dead-zone-skip configurations were not measured, and the floor's monotonicity in total shift records them as dead. Every configuration below the cropped axis range (total shift 15--33) diverged.
  (b) Contrast readout fidelity $F_c$ of the 657 configurations classified as converged.
  }
  \label{fig:FIG-S3}
\end{figure}

The qubit sweeps verify that discipline. We performed a 4824-configuration training sweep with the LMS filter on the transmon qubit denoted as "Q1", which mapped convergence over step size and total shift (Fig.~\ref{fig:FIG-S3}). 657 configurations converge in a narrow diagonal band. The best achieved $84.06\%$, a $+1.44\%$ selection maximum over the unfiltered baseline. That number was mechanism evidence rather than a deployable gain since the deployed-configuration of Sec.~\ref{sec:map} was a tie with the boxcar averager. The same sweep did not support the proposed cancellation-loss explanation, a fixed-point weight-quantization collapse distinct from the kernel collapse of Sec.~\ref{sec:governing}. If weight quantization were degrading the filter, fidelity should be worst where the weights were largest because large opposing weights have more cancellation loss. The data showed the opposite. Across the converged band, the binned mean fidelity increased monotonically with the coefficient energy, $|w|^2$ ($69.1\%$, $70.9\%$, $73.7\%$, $76.6\%$ across four log-spaced bins). The experimental operating points land inside that converged band, with $|w|^2 = 4.9\times10^{5}$--$1.5\times10^{6}$ and clean coefficient read-back.

The coefficient read-back is also where the kernel collapse stopped being an argument and became a measurement. The coefficients the fabric loaded were the tap stems of Fig.~\ref{fig:FIG-DT}(a), and their running sum accumulated to the collapse scalar with $|W| = 0.037$ at $\arg W = -0.242$~rad [exp.]. The bit-true model closed the arithmetic, reproducing the floating-point construction to a 99th-percentile residual of 36.7~LSB against the pre-registered 64-LSB bound (one truncation per tap). The residual reached the quantization floor after $m^{*} = 47$ samples were trimmed. We also compared decisions on held-out shots. After transforming the boxcar decision rule by $W$, the filtered and unfiltered classifiers assigned different labels to 398 of 50,000 shots ($0.80\%$), while the paired change in fidelity was near zero [offline].

\begin{figure*}[!t]
  \centering
  \includegraphics[width=\textwidth]{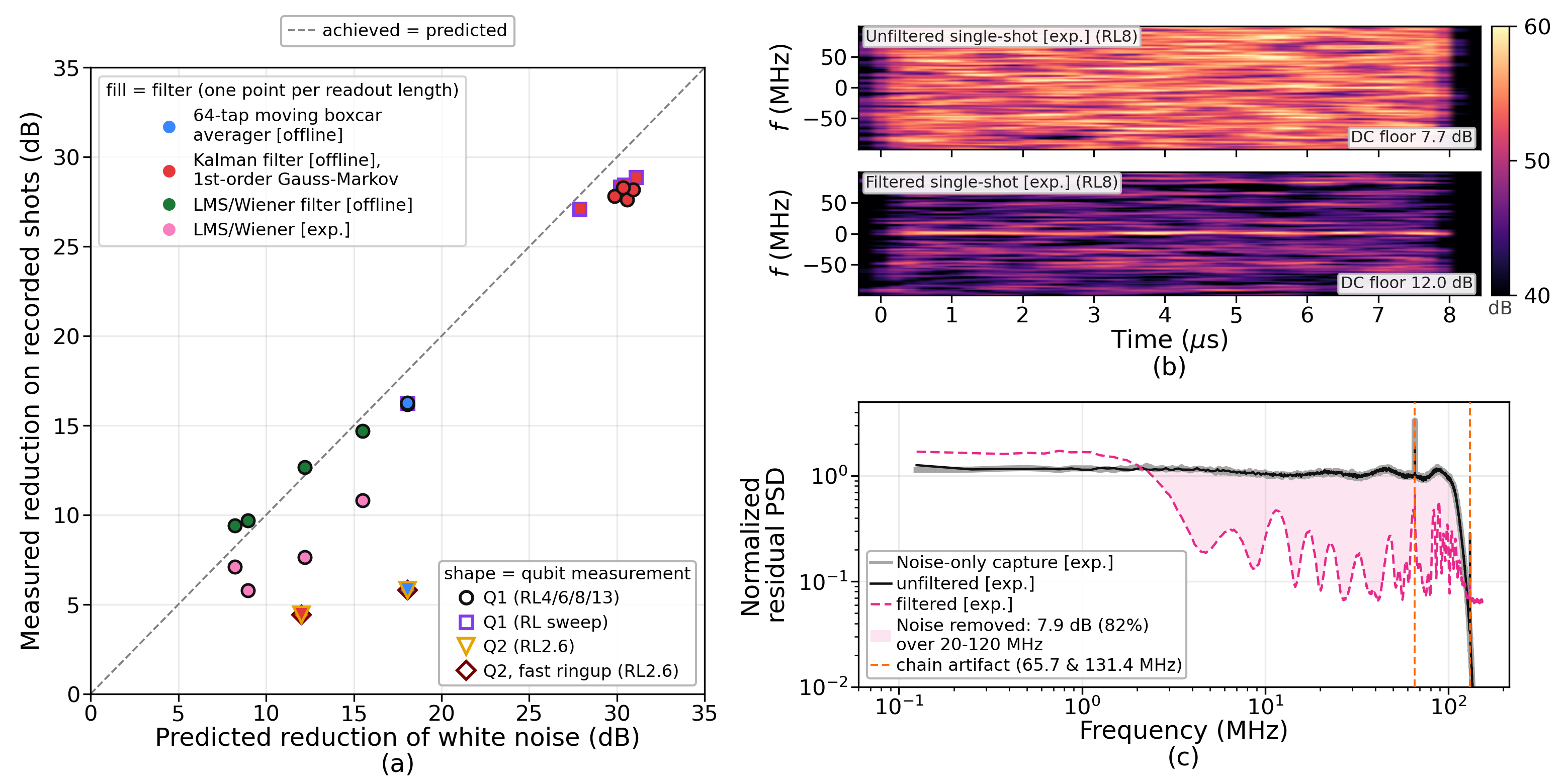}
  \caption{
  Predicted and measured noise reduction.
  (a) Noise reduction on measured shots compared with the white-noise prediction from Eq.~\ref{eq:inst:whitegain}. Color identifies the filter and whether it was applied in hardware or offline, while marker shape identifies the qubit acquisition. The smaller reduction on the ancilla (Q2) was consistent with its strongly correlated in-shot noise.
  (b) Single shots from unfiltered and filtered transmon acquisitions at RL~$= 8~\mu$s in Mode~1 [exp.]. The frequency axis gives complex-baseband offsets from the readout carrier $f_{\rm ro}$ after digital downconversion~\cite{Stefanazzi2022}. Positive and negative offsets correspond to RF frequencies above and below the carrier. Both rows use the same $[40, 60]$-dB display scale, smoothing, and $\pm100$~MHz range. The central line is unresolved at the analysis-window frequency resolution. Most of the measured state-dependent signal lay near DC, while the largest noise reduction occurred at higher offsets (Sec.~\ref{sec:governing-law}).
  (c) Pooled normalized residual spectra from the two RL~$8~\mu$s acquisitions in (b), shown over positive baseband frequencies [exp.]. The shaded region marks the 7.9-dB reduction over 20--120~MHz. The unfiltered residual agreed with a noise-only capture taken before the measurements, consistent with readout-chain noise dominating this band. The filtered spectrum lies above the unfiltered spectrum at the lowest plotted frequencies.
  }
  \label{fig:FIG-S6}
\end{figure*}

\subsection{Removable Noise for the Estimation Objective}
\label{sec:instrument-denoise-char}

The governing law section closed the increased fidelity claim. However, this does not discredit the filter removing a large, measurable amount of noise from a single shot time trace. That removal buys nothing for the state-assignment decisions, but a filter that follows the clean waveform is an observable for drift in the readout during qubit measurements. This tracking mechanism is posed at the end of this subsection and deferred to future work. 

Fig.~\ref{fig:FIG-S6} shows the noise removal from single-shot time traces in three panels. Fig.~\ref{fig:FIG-S6}(a) compares the reduction on measured shots replayed offline through algorithmically hardware-accurate filter models against the estimated reduction the same filter would achieve via Eq.~\ref{eq:inst:whitegain} on white noise. Fig.~\ref{fig:FIG-S6}(b) shows the experimentally measured noise reduction spectrogram for a single shot using the LMS/Wiener filter on the transmon qubit at RL~$= 8~\mu$s. Fig.~\ref{fig:FIG-S6}(c) shows the residual spectra of the same two acquisitions. Reductions are quoted over the full digitized band ($\pm153.6$~MHz), unless a band is named. 

Each recorded shot occupies $N$ complex dimensions. The signal itself spans essentially two dimensions (the class templates, $\mathbf{s}_g$ and $\mathbf{s}_e$), so the fraction of white-noise power lying off the signal is $1 - 2/N$, at least 99.3\% for every record here. The spectral form of the budget is the residual power outside the resonator bandwidth. This is integrated from the measured residual periodograms, and the residual is each shot minus its class-mean waveform. It is 23.0--33.1~dB across the measured Q1 readout lengths. 
The budget is also qubit-specific. For white noise it would be set by the ratio of the digitized band to the band kept around the resonator, which is 18.7~dB for the Q2 ancilla. The ancilla's in-shot residual is strongly colored and places only 5.7~dB out of band, and the Q1 residual is only mildly colored. 

Fig.~\ref{fig:FIG-S6}(a) plots the measured reduction in residual power against the reduction the same taps would give on white noise, which for a unit-DC-gain filter with weights $h[k]$ is
\begin{equation}
      R_{\rm pred} = 10\log_{10}\frac{\bigl|\sum_{k} h[k]\bigr|^{2}}{\sum_{k} \lvert h[k]\rvert^{2}}.
      \label{eq:inst:whitegain}
\end{equation}
The readout chain's amplifiers add broadband noise that fills the whole digitized band, and the unfiltered residual in Fig.~\ref{fig:FIG-S6}(c) matches a noise-only capture, so this band is chain noise and not the qubit. The resonator's response occupies only its 85-kHz linewidth near DC after downconversion. A single shot is a narrow signal under a wide sheet of noise, and nearly all of that noise lies where no signal is. An ensemble average across shots looks clean because the chain noise is independent from shot to shot. The filter does the same job inside one shot by averaging neighboring samples per pulse. The prediction assumes the noise is also independent from sample to sample, so the taps average it down while the signal, nearly constant across the 64-tap span, passes. Correlated noise does not average down, so a filter lands on the diagonal when its input noise is white over its span and falls below it by the share that is correlated. The Q1 offline replays sit on the diagonal, and the 64-tap average delivers 16.2--16.3~dB against 18.06~dB because the Q1 residual is only mildly correlated. The Q2 ancilla sits far below it because its residual is strongly correlated.

\begin{table*}[t]
	\caption{Measured Operating-Point Record}
	\label{tab:record}
	\centering
	\scriptsize
	\setlength{\tabcolsep}{1.5pt}
	\begin{tabular}{@{}ccccccccccc@{}}
	\toprule
\begin{tabular}[c]{@{}c@{}}Qubit\end{tabular} &
\begin{tabular}[c]{@{}c@{}}Operating\\Point\end{tabular} &
\begin{tabular}[c]{@{}c@{}}$T_1$\\($\mu$s)\end{tabular} &
\begin{tabular}[c]{@{}c@{}}$T_2^*$\\($\mu$s)\end{tabular} &
\begin{tabular}[c]{@{}c@{}}$\kappa/2\pi$\\(MHz)\end{tabular} &
$\delta/\sigma$ &
$\cos^2\theta$ &
\begin{tabular}[c]{@{}c@{}}Readout\\Fidelity\end{tabular} &
\begin{tabular}[c]{@{}c@{}}LMS/Wiener\\$\Delta$ (\%)\end{tabular} &
\begin{tabular}[c]{@{}c@{}}Per-sample MF\\$\Delta$ (\%) {[offline]}\end{tabular} &
\begin{tabular}[c]{@{}c@{}}Tuned Boxcar Avg.\\$\Delta$ (\%) {[offline]}\end{tabular} \\
	\midrule
	\begin{tabular}[c]{@{}c@{}}Q1, 2D transmon\end{tabular} &
RL~4~$\mu$s & 237.9 & 16.6 & 0.085 &
  1.33 & 0.50--0.76 & 50.8\% & $+0.29$ & $+3.00$ & $+5.56$ \\
 & RL~6~$\mu$s & & & & 2.09&  & 71.0\% & $-0.74$ & $+1.90$ & $+2.93$ \\
 & RL~8~$\mu$s & & & & 2.54&  & 79.8\% & $+1.17$ & $+2.62$ & $+2.43$ \\
 & RL~13~$\mu$s & & & & 2.68&  & 82.5\% & $+1.47$ & $+0.02$ & $+2.37$ \\
 & \begin{tabular}[c]{@{}c@{}}RL sweep (1--13~$\mu$s)\end{tabular} & 315.3 & 211.3 & &
  -- & $\sim$0.82 (RL~13) & \begin{tabular}[c]{@{}c@{}}78.26\%, 89.13\%$^{\rm ma}$\\(RL~13)\end{tabular} & -- & \begin{tabular}[c]{@{}c@{}}$+6.52$, $+3.26^{\rm ma}$\\(RL~4)\end{tabular} & -- \\
	\addlinespace
Q2, 3D ancilla & RL~2.6~$\mu$s & $\sim$165 & 36 & 2.07 & 4.17 & -- &
  $\sim$94.96\%, 97.48\%$^{\rm ma}$ & -- & $+0.50$ & -- \\
 & \begin{tabular}[c]{@{}c@{}}Fast ringup (2.6~$\mu$s)\end{tabular} & $\sim$165 & 36 & 2.07 & 4.05 & -- &
  94.08\%, 97.04\%$^{\rm ma}$ & -- & $+0.30$ & -- \\
 & \begin{tabular}[c]{@{}c@{}}RL sweep (0.2--10~$\mu$s)\end{tabular} & -- & -- & 2.07 & \begin{tabular}[c]{@{}c@{}}4.7 (RL~2.6)\end{tabular} & -- &
  \begin{tabular}[c]{@{}c@{}}94.4--96.2\% (RL $\geq 2.2$)\end{tabular} & -- & -- & -- \\
	\addlinespace
Q3, Flux-tunable & RL~4.8~$\mu$s & 66.2 & 16.9 & 0.88 & 3.44 & -- &
  91.45\% & -- & -- & -- \\
 & \begin{tabular}[c]{@{}c@{}}RL accrual (0.16--16~$\mu$s)\end{tabular} & 66.2 & 16.9 & 0.88 & \begin{tabular}[c]{@{}c@{}}3.87 (peak 6.9~$\mu$s)\end{tabular} & -- & 92.35\% (peak 6.9~$\mu$s) & -- & -- & -- \\
	\bottomrule
	\end{tabular}
	\par\vspace{2pt}
	{\raggedright\scriptsize All entries are the contrast convention; $^{\rm ma}$ gives the mean-assignment value, $F_{\rm ma} = (1 + F_c)/2$, beside the entries first reported in that convention.\par}
	{\raggedright\scriptsize [offline] columns are computed from the measured shots in post-analysis.\par}
\end{table*}

The Q1 hardware points also sit below the diagonal, and not because of the correlated noise. The FIR cores floor each convolution sum to an integer count before forming the 16-bit output. This adds a constant offset of about one count, and a flat quantization floor of 0.56~counts rms. The frozen kernel's DC gain is small ($|W| = 0.037$), so that floor is comparable to the residual the kernel leaves. The bit-true twin, carrying the same truncation, reproduces the fabric spectrum to 0.02~dB. Panel (b) shows the hardware removal on a single shot at RL~$= 8~\mu$s, where the floor drops beyond 10~MHz, and panel (c) shows the pooled residual falling 7.9~dB over 20--120~MHz [exp.]. The trained filter also falls short of the 64-tap average because noise in the trained taps widens its passband, from a two-sided ENBW of 4.8~MHz to about 39~MHz at RL~$= 8~\mu$s.

Even the 4.8-MHz passband is far wider than the 85-kHz linewidth, and leaves most of the budget in place. A Kalman filter is the optimal linear estimator for a resonator modeled as a first-order Gauss--Markov process~\cite[Ch.~10]{Haykin2002}. It predicts the resonator's slow response from its own past and keeps only what fits. On the measured shots, the unit-DC Kalman filter reached 27.1--29.4~dB, 0.8--3.0~dB below the budget at the chosen readout lengths. This sets the prediction model the firmware would be designed toward. Mixed-ensemble training costs the templates at most 0.11~dB, and the conditional-mean denoiser changes no decision. The filter therefore makes the resonator's response visible in one shot, an averaging the state assignment already gets from its sum.

\section{The Measured Operating-Point Record}
\label{sec:map}

The question of this paper is whether any on-fabric pro- cessing in the readout chain can beat conventional methods in dispersive readout. This section shows the measurement results, and defines where the identified methodologies of this research can reach effective gains.

\subsection{The Record and Its Disclosures}

The per-window deflection $\delta/\sigma$ is the accumulated state separation in units of the single-shot noise spread at the chosen readout window. Basically, this is a measurement for how far apart the center of mass for the qubit states are (such as in Fig.~\ref{fig:FIG-V1}), and is detailed in Eq.~\ref{eq:thA:deflection}. Its square is the discrimination SNR of Eq.~\ref{eq:thA:deflection}, $\mathrm{SNR}_{\rm disc} = (\delta/\sigma)^2$.  The deflection carries absolute meaning through the Gaussian-tail relation $F_c \approx 1 - 2Q(\delta/2\sigma)$, with $Q$ the Gaussian tail integral~\cite{Blais2021,Krantz2019,Gambetta2007}. The computed thresholds at $\delta/\sigma = 3.50$ (92\% contrast, 96\% mean-assignment) and 4.65 (98\%, 99\%) therefore mark where early decision becomes reachable.

The Q1 acquisitions used one fixed-frequency 2D transmon~\cite{Bal2023systematic} with a high-electron-mobility-transistor (HEMT)-amplified readout chain. An earlier acquisition swept the readout length (RL) from 1 to 13~$\mu$s at reduced drive. Its hardware LMS/Wiener results were excluded and replaced by the later acquisition reported here. The adaptive-filter firmware was operated only on Q1.

Q2 is a 3D-style transmon used as an ancilla qubit in the SRF-cavity platform~\cite{Roy2024Qudit}. It was measured on a separate testbed in the same refrigerator through a readout resonator near 7.49~GHz and a Josephson-parametric-amplifier (JPA)-amplified chain.

Q3 is a flux-tunable qubit on a nine-qubit device. Its setup used an OPX1000 controller rather than QICK, and a traveling-wave parametric amplifier (TWPA)-amplified chain was also employed. These calibration captures provided an additional fixed-window fidelity-accrual comparison in Sec.~\ref{sec:sprt}. These datasets were not used as controlled comparisons of qubit architecture or controller performance.

\begin{figure}[t]
  \centering
  \includegraphics[width=\columnwidth]{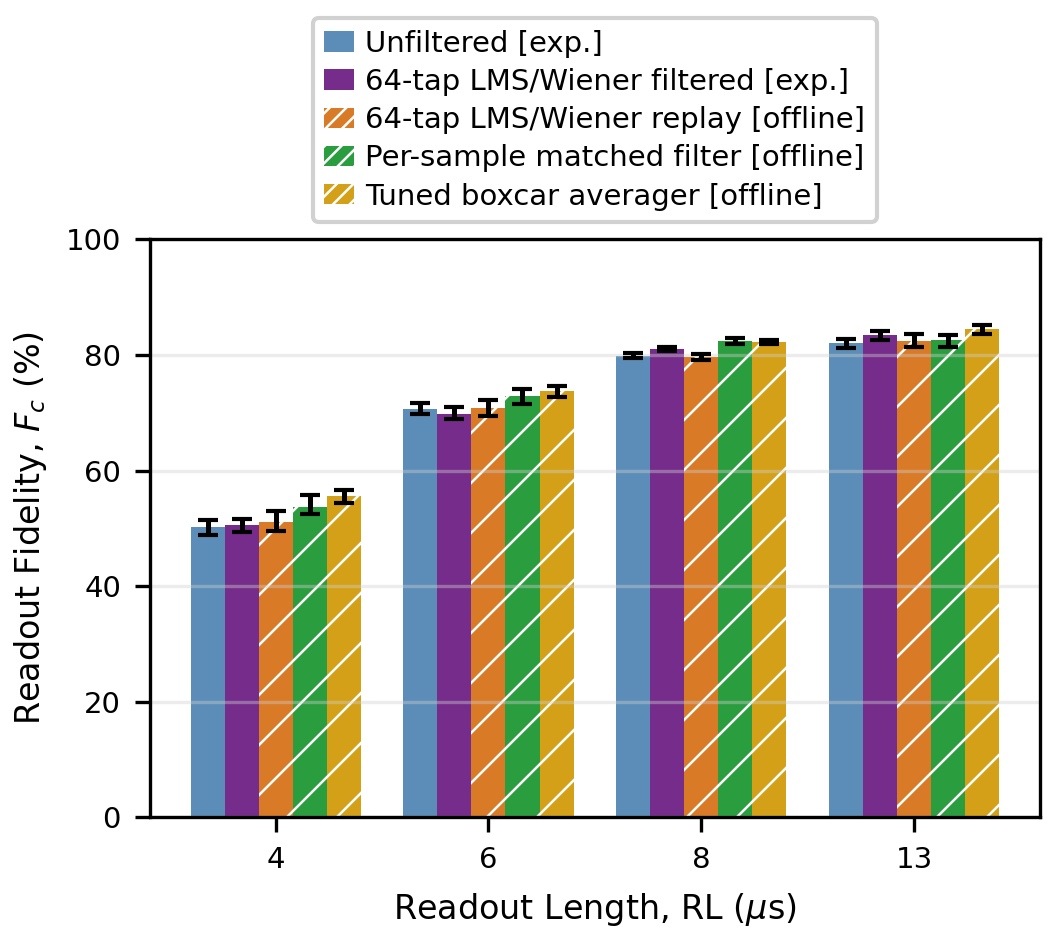}
  \caption{
  Readout methods at the four retained Q1 operating points. Solid bars show hardware acquisitions; hatched bars show offline analyses of recorded shots. Training and evaluation protocols are specified in the text. Same-shot frozen-filter replay shows no resolved fidelity difference from full-window boxcar integration. Window tuning accounts for much of the observed improvement. Whiskers indicate 95\% bootstrap confidence intervals.
  }
  \label{fig:FIG-A4}
\end{figure}

Table~\ref{tab:record} uses the fidelity conventions of Sec.~\ref{sec:instrument-model}. Offline comparisons used the recorded shots, and the experimental LMS/Wiener comparisons use separate filtered and unfiltered acquisitions. Fig.~\ref{fig:FIG-A4} shows the experimental and offline data. In Table~\ref{tab:record}, these are The Q1 boxcar and per-sample matched-filter entries are evaluated on the held-out test half, the hardware LMS/Wiener and Q3 entries use full captures, and the tuned-boxcar results use nested cross-validation. Coherence columns are same-acquisition bring-up fits, and the RL-sweep row's $\cos^2\theta$ is a whitened-template estimate. The per-sample MF column is the offline reference detector computed with knowledge of the class templates.

The retained acquisitions have RL $=4,6,8,$ and 13~$\mu$s. The remaining lengths were excluded because of an instrument fault, a Mode-0 artifact, or incomplete captures, except 2~$\mu$s, which was not acquired. The four retained lengths used readout amplitudes of 1900, 1450, 1150, and 780 DAC units, respectively. Their trends therefore reflect changes in operating point rather than readout length alone.

\subsection{The Frozen-Filter Tie and the Per-Sample Escape}
\begin{figure}[!t]
  \centering
  \includegraphics[width=\columnwidth]{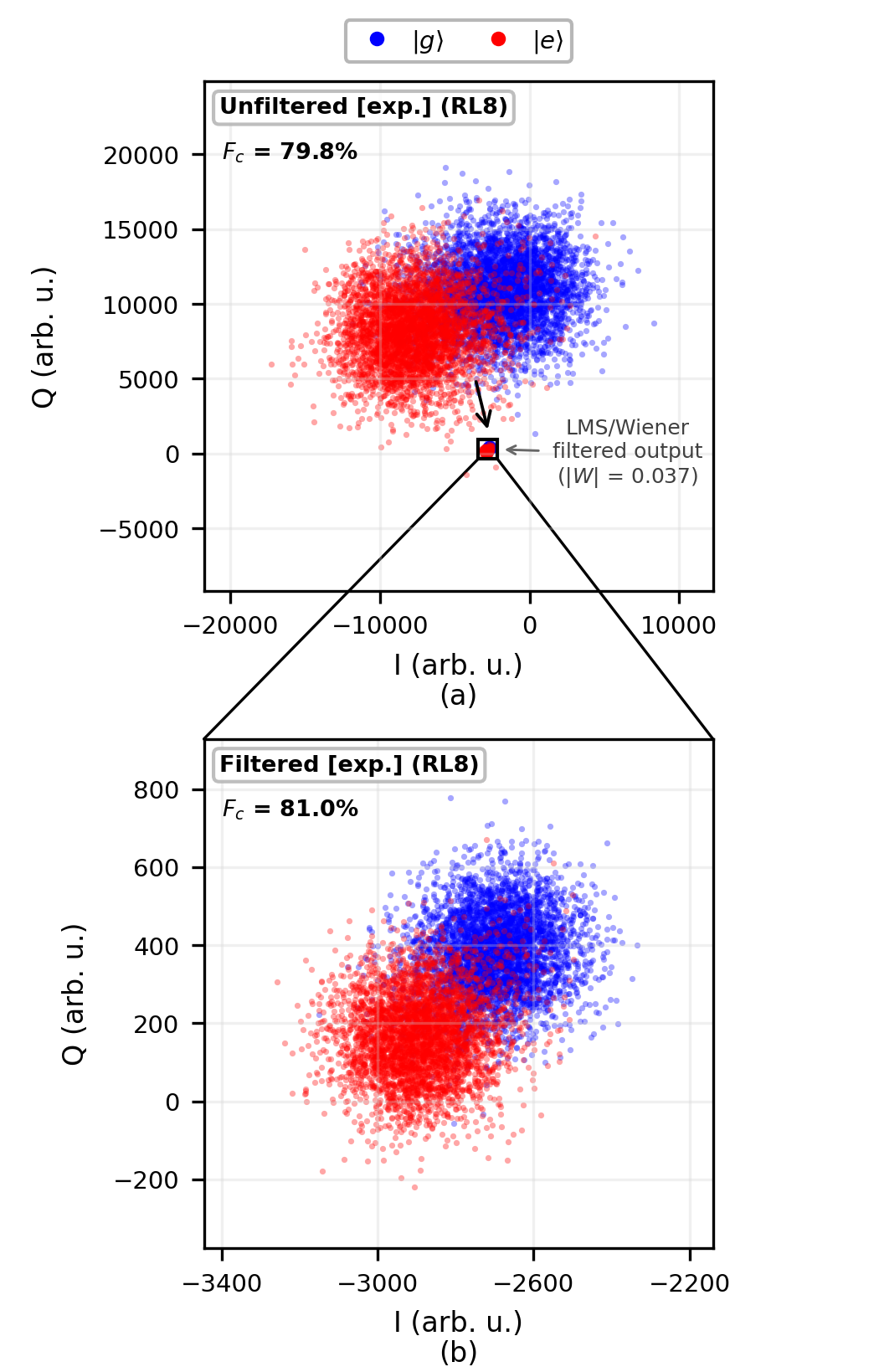}
  \caption{
  Separate unfiltered and filtered Q1 acquisitions at RL~$=8~\mu$s, with 50,000 shots per class [exp.]. (a) Both distributions on the unfiltered scale. (b) The filtered distribution enlarged. A common gain $|W|=0.037$ scales both the state separation and noise spread in the scalar approximation. The displayed fidelities, 79.8\% and 81.0\%, come from separate acquisitions and do not isolate the effect of filtering. Fig.~\ref{fig:FIG-A4} provides the same-shot comparison.
  }
  \label{fig:FIG-V1}
\end{figure}
\begin{figure}[!t]
  \centering
  \includegraphics[width=\columnwidth]{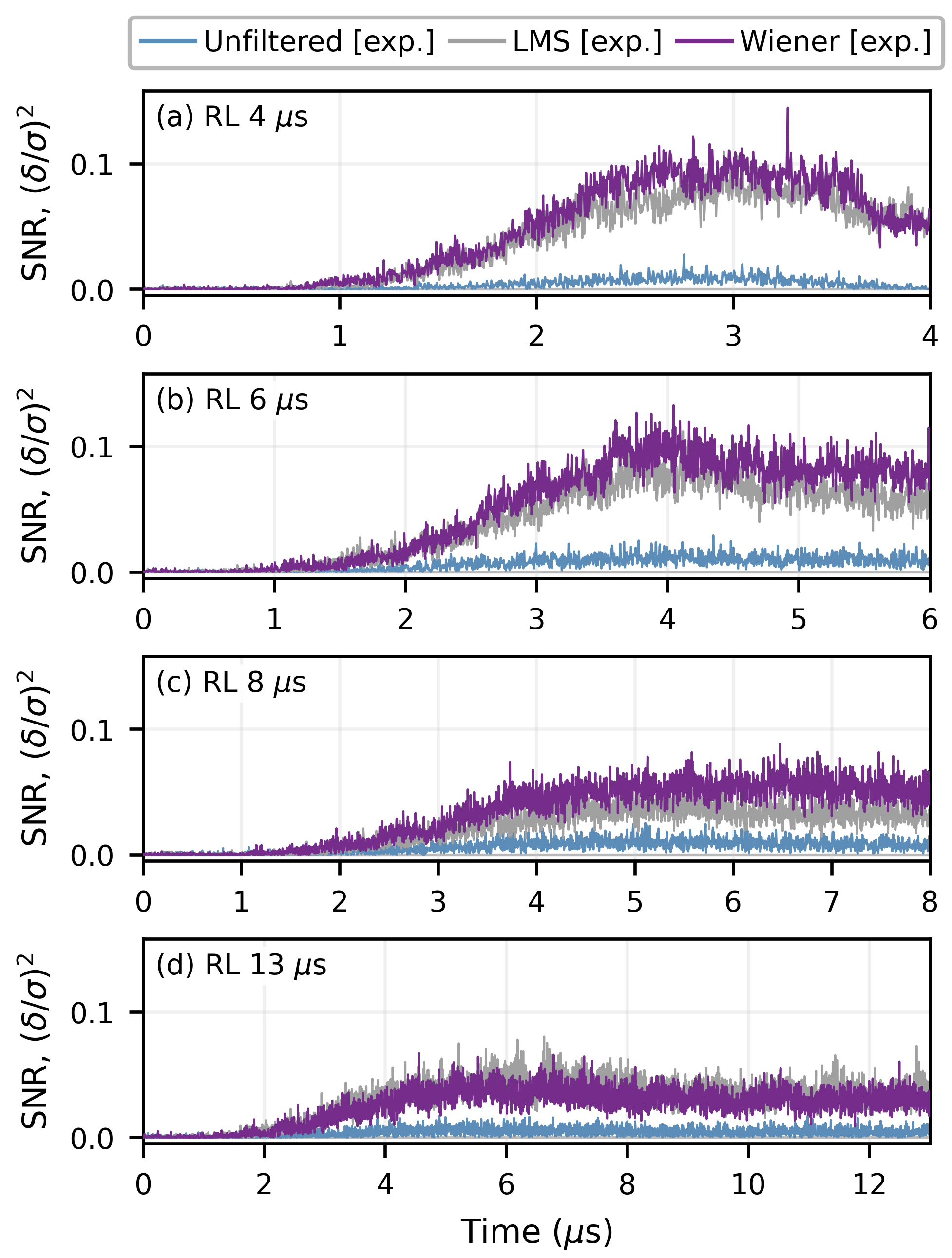}
  \caption{
  Per-sample SNR $(\delta/\sigma)^2$ against time at the four retained readout lengths on Q1 [exp.]. Each mode was acquired separately within one sitting, so level differences carry between-run drift as well as the filtering, and the same-shot control is the replay of Fig.~\ref{fig:FIG-A4}. The plotted statistic is the square of the per-sample deflection of Eq.~\ref{eq:thA:deflection}, each arm scored in its own train-fixed projection on the real axis (interleaved 50/50 split, 5,000 shots per class per half). Timing alignment and percentile comparisons are described in the text.
  }
  \label{fig:FIG-PS1}
\end{figure}

The 64-tap coefficients were trained on hardware, frozen, and replayed on the same shots used for the boxcar comparison. At all four retained readout lengths, no paired fidelity difference was resolved (Fig.~\ref{fig:FIG-A4}). Fig.~\ref{fig:FIG-V1} shows the I/Q cloud rescaling, but its separate acquisitions do not isolate filtering from timing differences or drift. Despite the increased per-sample SNR in Fig.~\ref{fig:FIG-PS1} and cleaner single-shot traces in Fig.~\ref{fig:FIG-S6}(b), the accumulated score remained approximately a rescaled boxcar statistic. The FIR mainly removes noise already attenuated by the boxcar (Fig.~\ref{fig:FIG-DT}(c)), but bandwidth alone is not the limitation. For frozen coefficients, uniform full-window accumulation gives $Y_{\rm FIR}\approx W X_{\rm total}$ apart from the finite-record edge (Eq.~\ref{eq:thA:collapse}). A nonzero $W$ scales the class separation and noise standard deviation equally, leaving $\delta/\sigma$ unchanged in the scalar limit. Retuning the passband does not remove this limitation while the approximation holds.

Fig.~\ref{fig:FIG-PS1} plots per-sample $(\delta/\sigma)^2$, whereas Table~\ref{tab:record} gives $\delta/\sigma$ for the full-window sum. The plotted SNRs do not add to the window SNR. Neighboring FIR outputs share input samples, so the variance of their sum includes covariance as well as individual variances.

Extending the FIR to as many taps as samples did not recover the proposed fidelity gain in the tested simulations. For the deployed filter, $M_{\rm eq}=0.9945$ at RL8 gives essentially unchanged shot requirements for independent repetitions of the same prepared state at equal accumulated discrimination SNR. The gains instead came from changing the decision weights. Tuned boxcar windows improved fidelity at all four lengths, while the matched filter added $+0.33\%$ beyond the tuned window under the common nested RL8 evaluation. These results identify the next architectural step: control how each sample contributes to the decision, rather than improve waveform reconstruction before the same accumulator. Sec.~\ref{sec:calibration} develops this direction through offline template-calibration and sequential-decision studies. They provide benchmarks for an on-fabric decision path whose readout performance and feedback latency must then be tested in situ.

\section{What Survives the Collapse}
\label{sec:calibration}

The preceding results motivate a decision path that uses the per-sample weights of Eq.~\ref{eq:thA:wgt} rather than a frozen FIR followed by uniform accumulation. These weights determine how each sample contributes to the state-assignment score.
A sequential rule then determines when to make the assignment. We examine two costs of this approach: the labeled shots needed to calibrate the decision templates and the time needed to assign an individual shot. Both are evaluated offline on Q2, with Q1 and Q3 providing operating-point comparisons.
The results provide calibration and decision-time benchmarks for an on-fabric implementation, rather than additional gains from the deployed FIR. Its readout performance and end-to-end feedback latency remain to be tested in hardware.

\subsection{Template Estimation from Fewer Labeled Shots}

\begin{figure}[t]
  \centering
  \includegraphics[width=\columnwidth]{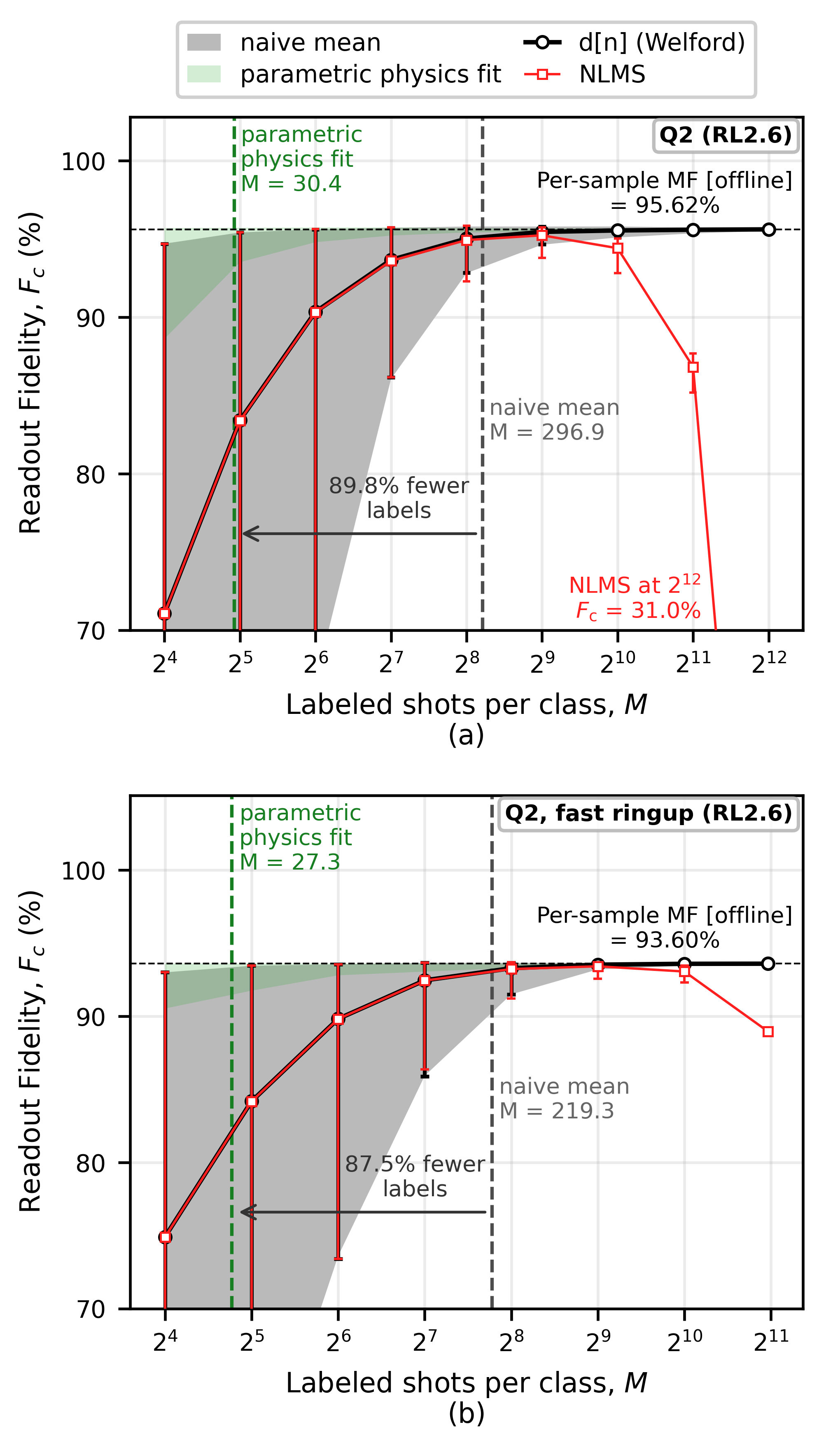}
  \caption{
  Template-calibration performance on Q2 [offline]. Contrast fidelity on a fixed test subset is plotted against labeled shots per class. Shading and error bars show central 95\% intervals over 200 resamples. Horizontal references use the full-training-subset matched filter. Panels (a) and (b) use constant-amplitude and fast-ringup readout pulses, respectively. Curves compare direct averaging, a parametric fit, and the $d[n]$ and NLMS templates described in the text. The off-scale NLMS point in (a) reflects arithmetic degradation in the offline twin, not the kernel-collapse identity.
  }
  \label{fig:FIG-S5}
\end{figure}

Matched-filter readout requires templates estimated from labeled calibration shots. Each template estimates the mean readout waveform for a prepared state. Fig.~\ref{fig:FIG-S5} shows how quickly four estimators of that template reach the reference fidelity as the number of labeled shots grows. Each estimator uses the same nested subsets of $M$ training shots per class, and the resulting classifiers are evaluated on a fixed test subset. The reference is the matched filter with templates estimated from the full training subset, and the target is a fidelity within 0.50\% of that reference. Here, $M$ counts calibration shots and not independent fidelity-validation shots.

The model-based estimator fits a ring-up waveform to each $M$-shot mean using variable projection~\cite{GolubPereyra1973}. Its linewidth grid spans 0.6--1.8 times an independent resonator-lineshape estimate, $\kappa/2\pi=2.33\pm0.43$~MHz, consistent with the 2.07-MHz ring-up value in Table~\ref{tab:record}. All parameters are fitted from the $M$ training shots; these counts exclude spectroscopy. For the constant pulse, about 30 shots per class reach the target versus 297 for direct averaging, an 89.8\% saving (89.2--89.8\% across three splits). Fast ringup requires about 27 versus 219 shots, an 87.5\% saving (83.7--88.6\% across four splits). The fitted linewidth specifies the template shape, not a physical linewidth measurement. Across the two tested grid centers, a factor-ten search span gives 75.5--79.2\% savings for the constant pulse and 60.6--68.5\% for fast ringup. The saving therefore depends on the allowed template family. The fast-ringup pulse adds a 0.20-$\mu$s pre-pulse at 2.9 times the subsequent amplitude, without shortening the acquisition window. Its matched-filter reference is 93.60\%, versus 95.62\% for the constant pulse. In a five-point pre-pulse-amplitude sweep at this duration, $P(e|g)$ rose from 2.0\% to 64\% [exp.], consistent with pre-pulse-induced $g\to e$ transitions.

The $d[n]$ curve uses the averager's integer update of Eq.~\ref{eq:inst:target}. The NLMS curve uses the offline twin's reference target, not its adaptive-FIR output. Its large-$M$ decline comes from floor division and does not establish the same failure in the production averager. Computed sequencer time for the template shots falls from $\sim$0.95~s for direct averaging to $\sim$0.097~s for the fit, excluding spectroscopy, software, and transfer overhead.
Under the Gaussian working model~\cite{VanTrees2001}, we compare the fidelity-estimate standard deviation over 200 resamples with the square root of the Cramer--Rao reference,
\begin{equation}
	\mathrm{CRB}(\widehat{F};M) = \frac{1}{4M}\Bigl[\phi(a_g)^2\bigl(1+\tfrac{1}{2}a_g^2\bigr) + \phi(a_e)^2\bigl(1+\tfrac{1}{2}a_e^2\bigr)\Bigr].
\label{eq:crb:floor}
\end{equation}
Here, $\phi$ is the standard normal density, and $a_g,a_e$ are the standardized distances of the class means from the threshold. Eq.~\ref{eq:crb:floor} applies to $F_{\rm ma}$; $F_c=2F_{\rm ma}-1$ makes its variance bound four times larger but leaves the efficiency ratio unchanged. The physics fit's $\mathrm{std}(\widehat{F})$ is $1.39\times$ the reference square root at $M=16$, and at or below it from $M=32$ onward on the full-rate dataset.

A three-way split uses separate subsets for training, target definition, and testing. The primary-acquisition saving changes from 89.8\% to 90.0\%. This tests template calibration, not a reduction in the independent shots needed to estimate a fixed classifier's fidelity.

\subsection{Sequential State Assignment}
\label{sec:sprt}

\begin{figure}[!t]
  \centering
  \includegraphics[width=\columnwidth]{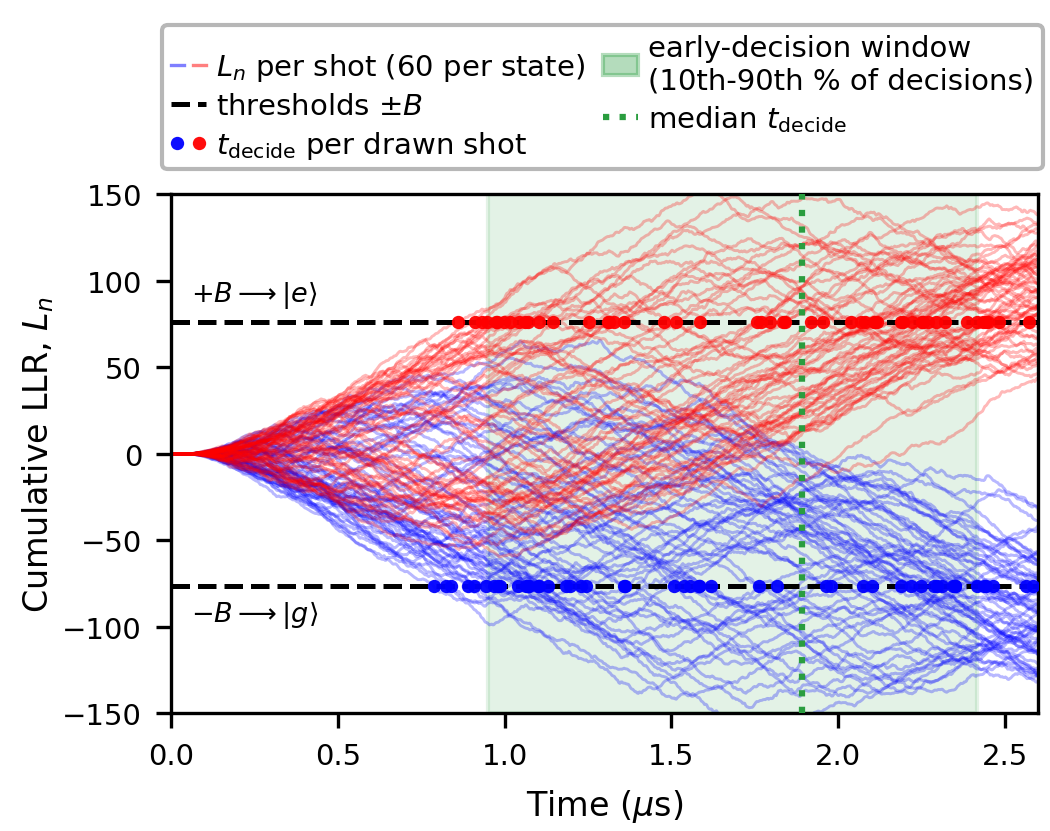}
  \caption{
  Sequential-discriminator replay on the Q2 2.6-$\mu$s capture [offline]. Thin curves show cumulative scores $L_n$ (Eq.~\ref{eq:sprt:llr}), and dots mark threshold crossings for the displayed shots. Shading gives the 10th--90th percentiles among shots crossing before the deadline. The dotted line is the median over all test shots, including terminal decisions at 2.6~$\mu$s. These are replayed decision times, not measured hardware latency.
  }
  \label{fig:FIG-A7a}
\end{figure}

Sec.~\ref{sec:map} examined decision weighting. Here, a Wald Sequential Probability Ratio Test (SPRT)~\cite{Wald1947} evaluates when to assign a shot; Fig.~\ref{fig:FIG-A7a} shows the Q2 replay. The Gaussian working model uses class means $\mathbf{s}_g,\mathbf{s}_e$ (Eq.~\ref{eq:thA:model}) and per-quadrature noise variance $\sigma_\eta^2[n]$, estimated on training data and held fixed. Its sample weight is
\begin{equation}
	w_{\rm sprt}[n] = \frac{s_e[n] - s_g[n]}{\sigma_\eta^2[n]},
\label{eq:sprt:w}
\end{equation}
the score increment is
\begin{equation}
	\ell[n] = \mathrm{Re}\Bigl\{ \conj{w_{\rm sprt}[n]}\,
	\bigl( x[n] - \tfrac{1}{2}(s_e[n] + s_g[n]) \bigr) \Bigr\},
\label{eq:sprt:inc}
\end{equation}
and the cumulative log-likelihood ratio (LLR) under the independent-noise model is
\begin{equation}
	L_n = \sum_{k \leq n} \ell[k] .
\label{eq:sprt:llr}
\end{equation}
The measured temporal correlations require empirical threshold calibration. The test stops at
\begin{equation}
	\tau = \min\!\left( \{\, 0 \leq n < N : |L_n| \geq B \,\} \cup \{N-1\} \right).
\label{eq:sprt:stop}
\end{equation}

The decision is the sign of $L_\tau$; shots without a crossing use the terminal score. $B$ is the smallest training-derived candidate whose training error does not exceed the boxcar error on the same training shots. It is held fixed during testing. Every reported replay time uses that threshold. Fig.~\ref{fig:FIG-A7a} shows threshold crossings and their timing distribution, not a measured hardware early stop.
Under the independent Gaussian model, the mean increment is positive for $e$ and negative for $g$, with magnitude half the sample score's squared deflection. For an $N$-sample window of length RL, Wald's expected-sample-number relation~\cite{Wald1947} gives
\begin{equation}
	\E[t_{\rm decide}] \approx \frac{2\ln\!\bigl((1-\beta)/\alpha\bigr)\,\mathrm{RL}}{(\delta/\sigma)_N^{2}}.
	\label{eq:sprt:esn}
\end{equation}
Here, $\alpha$ and $\beta$ are the two class-conditional error probabilities. For a uniform information rate, Eq.~\ref{eq:sprt:esn} predicts a mean decision within the window when $(\delta/\sigma)_N^2>2\ln((1-\beta)/\alpha)$, or $(\delta/\sigma)_N>2.43$ at the $(0.05,0.05)$ budget. This approximation does not guarantee that a capped detector meets that budget; the replay evaluates errors and timing together.

On the primary split of the Q2 2.60-$\mu$s capture (Fig.~\ref{fig:FIG-A7a}), 85.0\% of test shots crossed before the deadline. Including terminal decisions at the window end, mean decision time was 1.80~$\mu$s (69.2\% of the window) and median time was 1.89~$\mu$s. Contrast fidelity was 95.34\% versus 95.20\% for the boxcar, a paired difference of $+0.14$ percentage points (95\% CI $[-0.18,+0.46]$).
We assessed equivalence retrospectively using a $\pm0.5$-percentage-point margin in contrast fidelity, matching the calibration tolerance. The margin was fixed before the three-seed sweep but after the exploratory primary result. Equivalence required the paired 90\% bootstrap CI to lie within the margin; the primary CI $[-0.12,+0.40]$ met this criterion.
Across three overlapping splits of the same capture, mean times were 1.69--1.80~$\mu$s (65.0--69.2\% of the window), with 85.0--89.8\% in-window decisions. One sensitivity split also met equivalence; the other gave $-0.48$ percentage points (95\% CI $[-0.86,-0.10]$), a resolved decrease without established equivalence. Thus, timing improved on all splits, but equivalence was not established on all splits. These are sensitivity checks, not independent replications. Hardware feedback must also account for processing latency and the decision-time tail.

The transmon (Q1) does not reach this regime at its measured operating points. A $(0.05, 0.05)$ error budget needs $\delta/\sigma \geq 3.29$ at the window end, which the Gaussian relation of Sec.~\ref{sec:map} places at 90\% contrast fidelity (95\% mean-assignment). The ancilla enters at 4.17 and decides early, Q3 enters at 3.44 and sits just above the floor, and the transmon's 1.33--2.68 falls short. Therefore, the Q1 analysis does not establish early readout at that target. The limit belongs to the measured operating points and the detectors evaluated, not to planar qubits or to stopping rules in general. The floor is necessary rather than sufficient because the colored noise and the non-uniform information rate of a real record set where the score actually crosses, which is why Fig.~\ref{fig:FIG-A7a} reports the replay rather than the formula.

\begin{figure}[!t]
  \centering
  \includegraphics[width=\columnwidth]{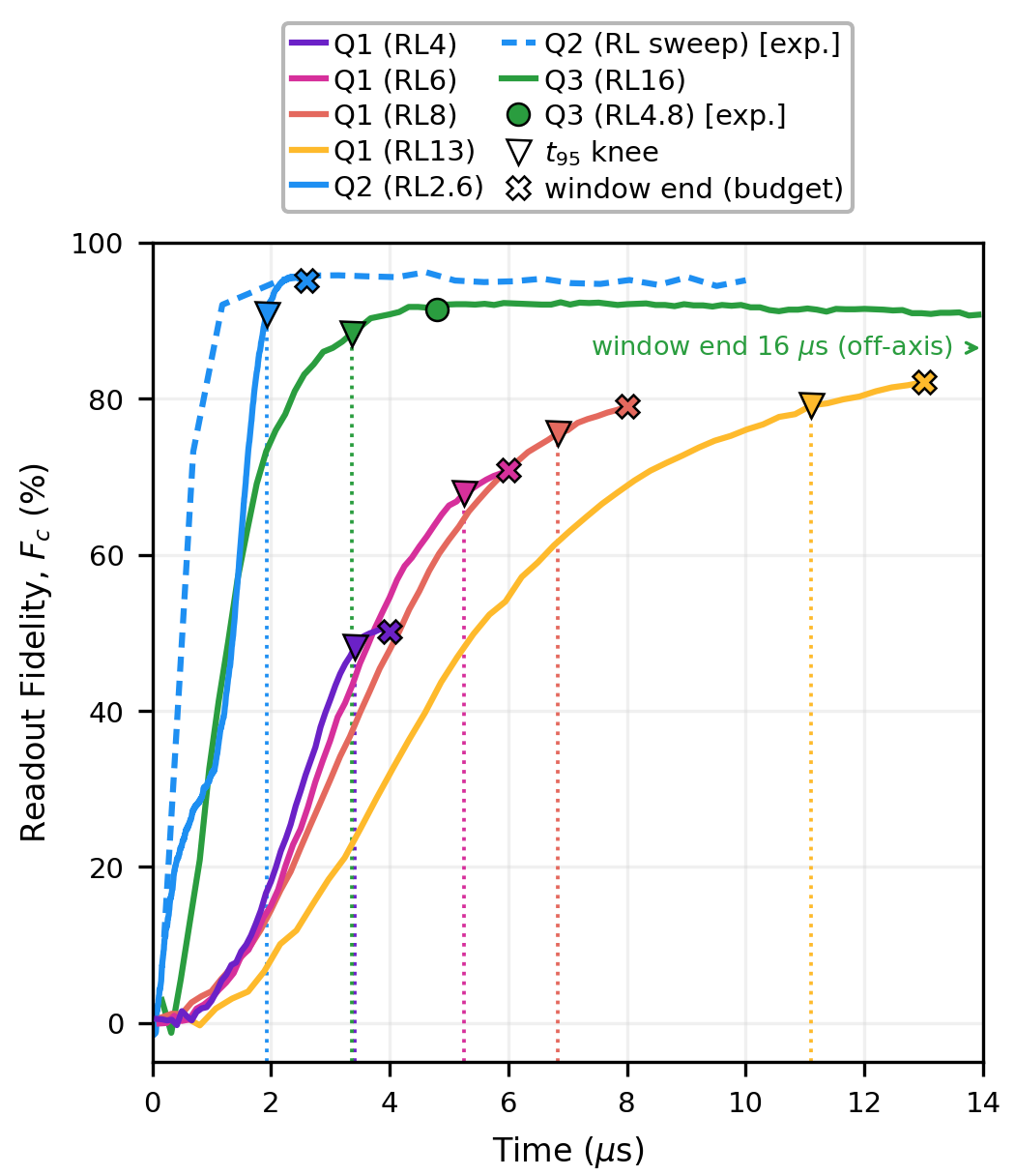}
  \caption{
  Fixed-window contrast fidelity for Q1, Q2, and Q3. Solid curves are prefix analyses of recorded traces. The dashed blue curve shows a separate Q2 hardware window sweep from an earlier sitting without the narrowband interferer, 3000 shots per state at each window length [exp.], and the filled green point shows Q3's 4.8-$\mu$s acquisition [exp.]. Triangles mark the first duration reaching 95\% of each curve's own peak, and crosses mark acquisition endpoints. Q3's 16-$\mu$s endpoint lies outside the displayed range. No sequential stopping rule is applied. The Q2 acquisitions and their relationship are described in the text.
  }
  \label{fig:FIG-A7b}
\end{figure}

The tested Q1 amplitude sweep did not reach the target. Moving the transmon into the early-decision regime would therefore require a change of operating point or readout chain. Q3 provides an intermediate fixed-window reference. At RL $=4.8~\mu$s its measured contrast fidelity is 91.45\%, and Fig.~\ref{fig:FIG-A7b} shows its fidelity accrual over the longer record. This fixed-window result alone does not establish whether a sequential rule would reach the same fidelity earlier. No sequential-detection result is claimed for Q3.

Fig.~\ref{fig:FIG-A7b} shows how fidelity accrues with integration time for the three qubits, with no stopping rule applied. The marked $t_{95}$ is the first duration reaching 95\% of each curve's own peak, and the cross is the window the acquisition used. Q2 flattens well inside its window, so the stopping rule of Fig.~\ref{fig:FIG-A7a} has a plateau to stop on. Q3 flattens before its 4.8-$\mu$s hardware point and sits on that plateau. The Q1 curves are still rising at their acquisition endpoints and do not establish early decisions at the $(0.05,0.05)$ target. The lower-target replay is discussed separately below. The $t_{95}$ marker is a descriptive reference, and not necessarily the fixed window that matches the sequential detector's fidelity. A comparison intended to isolate the benefit of per-shot stopping should select the fixed-window baseline at the same error or fidelity target.

Q2 reached the 90\% floor at different window lengths in different acquisitions. The 2.6-$\mu$s capture contained a narrowband residual near 0.43~MHz. A boxcar window strongly attenuates a stable tone near an integration duration of $1/f \approx 2.3~\mu$s, consistent with the observed fidelity peak near 2.4~$\mu$s. Offline tone removal moved the floor crossing to a shorter window. Its relevance to real-time stopping depends on whether the tone estimate can be obtained without later samples. A second hardware sweep, taken seven minutes after the capture in the same sitting and not drawn in Fig.~\ref{fig:FIG-A7b}, agrees with the capture's prefix curve to 1.3\% over 1.3--2.6~$\mu$s. Cutting a recorded trace short reproduces a real shorter window, and the slower rise belongs to that sitting.

For Q1, with the threshold set to the boxcar's own error rate and not the $(0.05, 0.05)$ budget, the replay's mean decision time was 73--81\% of the window. That is a different requirement and not an early-decision result at the registered target. At the tested Q2 operating points, delaying the integration start improved short-window fidelity, while windows longer than 2.2~$\mu$s changed by at most 2\%. Together with the tuned windows of Sec.~\ref{sec:map}, this supports evaluating window placement during calibration.

Decision timing depends on both the time-dependent state separation and the correlations between samples.
The replay therefore evaluates the accumulated score directly rather than relying only on Eq.~\ref{eq:sprt:esn}.

The FPGA design includes a sequential-decision engine with AXI-readable status and a 1024-sample window capacity. It has met timing constraints, but its readout performance has not yet been measured in hardware. The replay provides targets for testing classification fidelity, decision timing, and end-to-end control latency.

\section{Conclusion and Outlook}
\label{sec:conclusion}

This work asked whether additional digital processing improves superconducting-qubit readout beyond full-window and tuned boxcar integration. The on-fabric filter removes a measured amount of noise from every single-shot trace as the conjecture expected. The step from a cleaner trace to a better state assignment did not follow. For a frozen FIR followed by uniform accumulation, the accumulated output is the boxcar statistic times one complex number, apart from finite-record and arithmetic corrections. The coefficients and whatever trained them set that number, and it scales the state separation and the noise alike. This is what we call the boxcar limit, outlined in Sec.~\ref{sec:governing}.

Same-shot replay of the measured 64-tap LMS/Wiener coefficients showed no resolved fidelity improvement over full-window boxcar integration at the four retained Q1 operating points. The equivalent-shot ratios likewise showed no systematic reduction in independent repetitions at equal accumulated discrimination SNR. Window placement and per-sample weighting produced the gains that the frozen filter did not. A boxcar averager with window start and length chosen on training data beat full-window integration at every retained readout length [offline]. At RL~$=8~\mu$s, the per-sample matched filter added $+0.33\%$ beyond the tuned window under the common nested evaluation.

With a linewidth search range set from an independent resonator measurement, model-based template estimation on Q2 required 89.8\% fewer labeled template shots than direct averaging. Offline sequential replay on the primary split gave a mean decision time of 1.80~$\mu$s (69.2\% of the 2.60-$\mu$s window), counting terminal decisions at the window end. Across three overlapping splits of the same capture, mean decision times ranged from 1.69 to 1.80~$\mu$s. The primary split and one sensitivity split met the 0.5-percentage-point fidelity-equivalence criterion; the remaining split showed a resolved decrease. These results identify a shorter decision-time regime while exposing the fidelity constraint that hardware validation must address.

The deployed instrument tested the frozen-filter result on the fabric and exposed its finite-precision update effects. Coefficient readback and the bit-true model linked the measured behavior to the implemented arithmetic. The same telemetry is a candidate for detecting readout drift, but its value for triggering recalibration remains to be tested.

The boxcar limit applies to the tested frozen-FIR-plus-accumulator path, not to linear readout processing generally. The results identify the next architectural step: direct control of the decision weights and the time of assignment. Future work will test this path in situ for state assignment and active-reset feedback, and evaluate drift-triggered recalibration. For FPGA-based quantum controllers~\cite{Stefanazzi2022,Vora2024QubiCML,DiGuglielmo2025}, this work establishes a measured limitation of the tested processing path and provides offline calibration and timing benchmarks for the architecture that follows.

\section*{Acknowledgment}

This material is based upon work supported by the U.S. Department of Energy,
Office of Science, National Quantum Information Science Research Centers,
Superconducting Quantum Materials and Systems (SQMS) Center, under Contract
No.~89243024CSC000002. Fermilab is operated by Fermi Forward Discovery Group,
LLC, under Contract No.~89243024CSC000002 with the U.S. Department of Energy,
Office of Science, Office of High Energy Physics.

\bibliographystyle{IEEEtranNhref}
\bibliography{lms_v1_TQE26}

\end{document}